\documentstyle[psfig]{cupconf}

\ifoldfss
\else
  \ifnfssone
    \newmathalphabet{\mathit}
      \addtoversion{normal}{\mathit}{cmr}{m}{it}
      \addtoversion{bold}{\mathit}{cmr}{bx}{it}
    \newmathalphabet{\mathcal}
      \addtoversion{normal}{\mathcal}{cmsy}{m}{n}
    \else
    \ifnfsstwo
    \fi
  \fi
\fi

\def\hexnumber#1{\ifcase#1 0\or1\or2\or3\or4\or5\or6\or7\or8\or9\or
 A\or B\or C\or D\or E\or F\fi }

\makeatletter
\ifx\CUP@mtlplain@loaded\undefined
\else
\fi
\makeatother
 \makeatletter
 \ifx\CUP@mtlplain@loaded\undefined
   \font\tenbmi=cmmib10 at 10pt
   \font\sevenbmi=cmmib10 at 7pt
   \font\fivebmi=cmmib10 at 5pt

   \newfam\bmifam
   \textfont\bmifam=\tenbmi
   \scriptfont\bmifam=\sevenbmi
   \scriptscriptfont\bmifam=\fivebmi
   
 \fi
 \makeatother
\ifnfsstwo

\fi
\ifnfssone

\fi
\ifoldfss

\fi

\mathchardef\varLambda="0103

\makeatletter
\ifx\CUP@mtlplain@loaded\undefined
\else
\fi
\makeatother
\makeatletter
\ifx\CUP@mtlplain@loaded\undefined
  \font\tenbms=cmbsy10
  \font\sevenbms=cmbsy10 at 7pt
  \font\fivebms=cmbsy10 at 5pt
  \newfam\bmsfam
  \textfont\bmsfam=\tenbms
  \scriptfont\bmsfam=\sevenbms
  \scriptscriptfont\bmsfam=\fivebms

  \edef\bsy@{\hexnumber\bmsfam}
  \mathchardef\bnabla="0\bsy@72
\fi
\makeatother
\title[Quasar Absorption Lines]{Quasar Absorption Lines}

\author[Jill Bechtold]%
{J\ls I\ls L\ls L\ns B\ls E\ls C\ls H\ls T\ls O\ls L\ls D\ls $^1$}

\affiliation{$^1$Steward Observatory and the Department of Astronomy, the University of Arizona,
Tucson, AZ 85721, USA\\[\affilskip]
}

\begin{document}
\ifnfssone
\else
  \ifnfsstwo
  \else
    \ifoldfss
      \let\mathcal\cal
      \let\mathrm\rm
      \let\mathsf\sf
    \fi
  \fi
\fi

\maketitle

\begin{abstract}
\noindent
$``$Analyzing a spectrum is exactly like doing a crossword puzzle, but
when you get through with it, you call the answer research."

{\it -- Henry Norris Russell.}

\bigskip

The absorption lines observed in quasar spectra have given us a 
detailed picture of the
intergalactic medium and the metal abundance and kinematics of high redshift
galaxies.  In this review, we present an introduction to the field, starting
with the techniques used for 
interpreting absorption line spectra.  We then survey the observational
and theoretical development of our understanding of the Lyman $\alpha$ forest,
the metal absorbers, and the damped Ly$\alpha$ absorbers.  We conclude with a
discussion of some of the remaining outstanding issues, and 
prospects for the future. 
\end{abstract}

\firstsection 

\section{Introduction}

Absorption lines were seen in the earliest photographic
spectra of quasars in the mid-1960's 
(Sandage 1965, Gunn $\&$ Peterson 1965,
Kinman 1966, Burbidge, Lynds $\&$ Burbidge 1966, Burbidge 1967).
By 1969, the concensus was that quasar redshifts 
are cosmological, and that many of the absorption lines in quasar
spectra originate in intervening galaxies (see Bahcall \&
Salpeter 1965, Bahcall \& Spitzer 1969).   Subsequently, 
W. L. W.  Sargent, Peter J. Young and collaborators wrote a series of
papers presenting the first comprehensive studies of QSO 
absorption lines, using high quality spectra 
obtained with Boksenberg's photon counter, the IPCS, 
at the Double Spectrograph on the Palomar Hale telescope
(Young et al. 1979; Sargent et al. 1979; 
Sargent, Young, Boksenberg \& Tytler 1980;
Young, Sargent \& Boksenberg 1982ab; Sargent, Young \& Schneider 1982). 

A classification scheme was developed, 
summarized in an article in the Annual Review of Astronomy
and Astrophysics, by Weymann, Carswell \& Smith (1981). 
First, the so-called $``$metal-line absorbers"  with redshifts,
$z_{abs}$, much less than the quasar emission line redshift, 
$z_{em}$, were 
attributed to interstellar gas in intervening galaxies.
Systems which were detected in Ly$\alpha$ and occasionally
the Lyman series of hydrogen, were dubbed the Ly$\alpha$ forest.
These showed no clustering like galaxies, nor detectable
metals.  They were thought to be primordial, pre-galactic,
inter-galactic gas, pressure-confined by a hypothesized 
inter-galactic medium, and ionized by the integrated UV radiation     
of quasars themselves (Sargent, Young, Boksenberg \& Tytler 1980).  
Finally, some quasars showed broad,
P-Cygni like troughs of absorption at the emission line redshift;
these were interpreted as radiatively driven winds associated with  
the quasar central engine (Turnshek 1984 and references therein).

Quasar absorbers are useful probes of the high redshift universe
for many reasons.  Atoms have a rich ultraviolet absorption 
spectrum; typically a single spectrum will contain lines
from many elements in a range of ionization states.  
Lines like Ly$\alpha$ are extremely sensitive probes 
of small amounts of gas.  Absorption spectra are relatively
easy to interpret, since the lines arise from atoms in the ground
state, and probe a pencil
beam through the volume of gas.  By contrast, emission lines
are a complicated integral of density and other physical 
conditions over the emitting volume.  Until recently, 
quasar absorbers contained the only information about $``$normal",
that is, non-active galaxies at high redshift.  They still are the best 
way  to study the detailed kinematics and metal abundances of high redshift
galaxies.   

Other reviews of the subject can be found 
in the conference proceedings edited
by Meylan (1995) and Petitjean $\&$ Charlot (1997).  
The Ly$\alpha$ forest was reviewed by Rauch (1998) 
and damped Ly$\alpha$ absorber abundunces by 
Lauroesch et al (1996).
 
\section{Analysis of Absorption Line Spectra}

\begin{figure}
\centerline{\psfig{figure=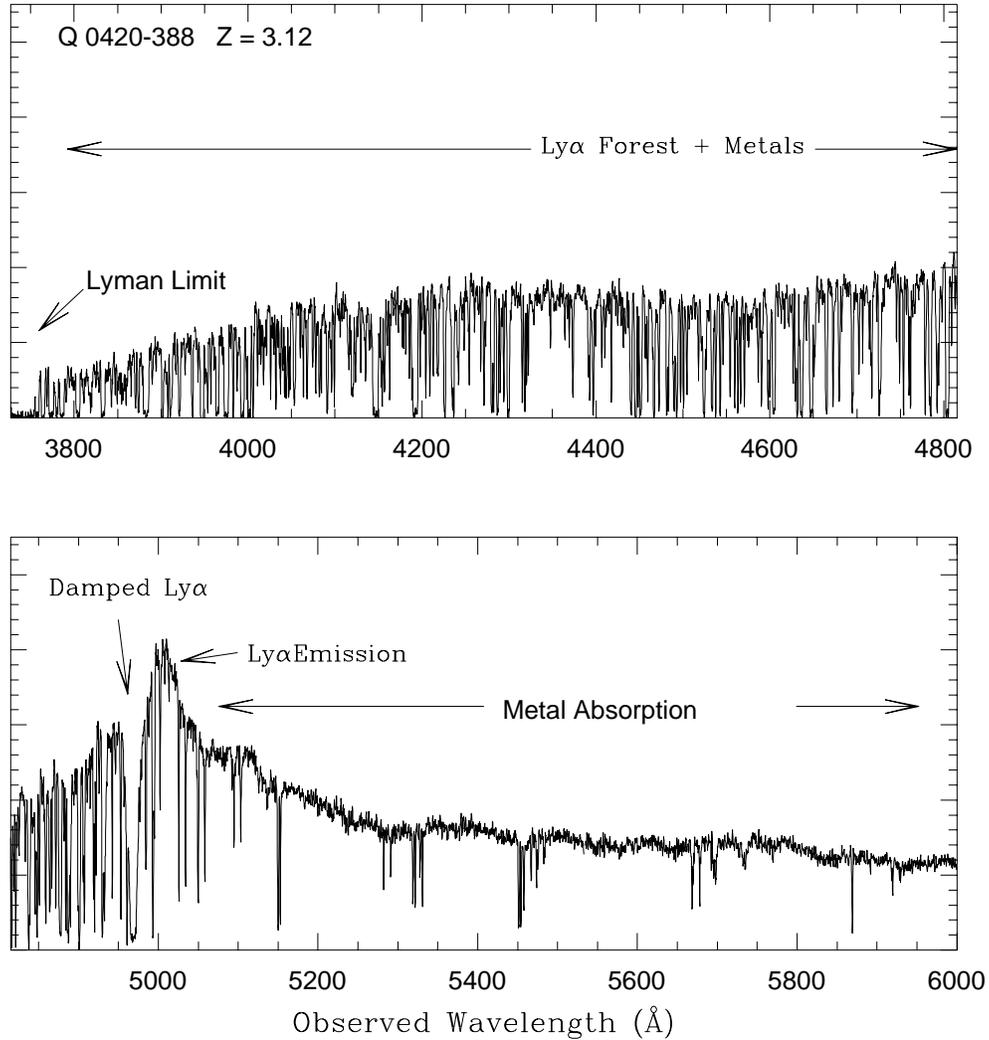,width=15cm,angle=0}}
\caption{High resolution spectrum of a $z=3.12$ quasar obtained with the
Las Campanas echelle spectrograph, by the author and S. A. Shectman.
Prominent features are indicated.}
\end{figure}

The spectrum of a z$\sim$3 quasar is shown in Figure 1.  The broad emission
line at $\lambda_{obs}=5000$ \AA \ is the Ly$\alpha$ emission 
associated with the quasar. The numerous absorption lines
blueward the emission line are mostly Ly$\alpha$ lines at 
different redshifts $z_{abs} < z_{em}$, where 
\begin{equation}
\lambda_{obs} = (1+z_{abs}) \lambda_{rest} 
\end{equation}
\noindent
and $\lambda_{rest}$= 1215.67 \AA \ for Ly$\alpha$.  
The broad feature just short of 5000 \AA\ is a damped Ly$\alpha$
line arising from an intervening absorber, which also causes the
observed continuum of the quasar to go to zero at the Lyman limit. 
The lines redward of the Ly$\alpha$ emission line are metal 
transitions associated
with the damped Ly$\alpha$ absorber as well as other redshift systems
along the line of sight.

Comprehensive lists 
of lines commonly seen in quasar spectra can be found in 
Morton (1991, 2000) and Verner, Barthel \& Tytler (1994).
Morton, York \& Jenkins (1988) has a short list of the strongest
lines suitable for many applications.

Spectra for wavelengths $\lambda_{obs} > 3200\ \AA$ are observable 
with ground-based telescopes,
whereas spectra at shorter wavelengths must be obtained with 
telescopes above the atmosphere.  On the red end of the optical, 
CCDs become transparent
to photons at $\sim9000$~\AA\ and although a few cases of absorption features
have been reported in the near-IR, (e.g. Elston et al. 1996) 
the sky background in the IR makes observing faint quasars difficult from
the ground.  In the rocket UV, the {\it Hubble Space Telescope (HST)} and 
the {\it International Ultraviolet Explorer (IUE)} 
have been used for spectroscopy in the wavelength range $1150-3200$ \AA.
The short wavelength limit arises from the cutoff in the MgFl coatings
for the optics.  Coatings for shorter wavelengths have relatively low 
reflectivity, so specialized missions have been launched for
spectroscopy in the $911-1150$\ \AA \ region -- most recently, the
{\it Hopkins Ultraviolet Telescope (HUT)} and the {\it Far Ultraviolet
Spectroscopic Explorer (FUSE)}.  The Milky Way
is opaque between 911 \AA\ and 62 \AA\ or 0.2 $keV$\ in the X-rays.
The {\it Chandra X-ray Observatory} and {\it XMM-Newton} are being used to study quasar absorption lines
in the X-rays (Bechtold et al. 2001), but results were not available in time 
for inclusion in this review. 
 
\subsection{Voigt Profile Fitting, the Curve of Growth and the Doublet Ratio}

How do we derive physical quantities of interest from the observed
absorption line spectra?   We summarize the basics in this section.
The results are derived in full in a number
of introductory texts, for example Gray (1992), Spitzer (1978), and
Swihart(1976).

\subsubsection{Voigt profile fits}

If the spectral resolution of the observation is good enough that the intrinsic
line widths are resolved, Voigt profiles can be fit to individual line
profiles to derive properties of the absorbing gas.  
The observed line profile is the line
profile for an individual atom convolved with (1) a function describing the 
broadening due to the distribution of 
atomic velocities (generally assumed to be Maxwellian with temperature $T$), 
and (2) the instrumental profile (generally modeled as a Gaussian).  
In addition, one expects
that for a sight line passing through a galaxy, 
the absorption feature will be produced by 
several $``$clouds" or $``$velocity components", 
each with its own temperature $T$, and some overall velocity  
width characteristic of the kinematics of the galaxy, sometimes
called the turbulent velocity.  
When the velocity dispersion 
of the clouds is comparable or smaller than the 
thermal widths of individual clouds, the individual velocity components 
will be blended.  

The absorption line profile for an individual atom, called the 
natural line profile,
is characterized by a cross-section which is a Lorentzian function of frequency $\nu$ 
\begin{equation}
\sigma_{\nu} = {{\pi e^2} \over {m_e  c}} \  f \ { {\Gamma/4\pi^2} \over 
{ (\nu - \nu_o)^2 + (\Gamma/4 \pi )^2}}  \ cm^2
\end{equation}
\noindent
where $f$ is  the oscillator strength, $\Gamma$ is the damping parameter,
and $\nu_0$ is the central frequency of the transition.  The term $``$damping"
arises from the semi-classical derivation of Equation 2.2 from consideration
of a harmonic oscillator with a force that damps the particle oscillations 
arising from the inevitable emission
of radiation by accelerating charges (e.g. Rybicki \& Lightman 1979).
The oscillator strength, or $f$-value, measures the quantum mechanical departure 
from the classical harmonic oscillator.  

For the transitions of interest, the typical width of the Lorenzian is tiny,  
$\Delta \lambda \sim 10^{-4} \AA$ 
full-width-half-maximum (FWHM) at $\nu_{0}$=10$^{14}$ Hz, 
or 0.006 km sec$^{-1}$ at 5000 \AA.  Thus the natural line width is much 
smaller than the thermal width of the absorbing Maxwellian velocity 
distribution of the atoms in the absorbing cloud, the instrumental 
line width,  or the turbulent cloud velocity dispersion.

The convolution of the Lorentzian profile for an individual atom 
with a Maxwellian results in 
a cross-section for absorption described by the Voigt integral 

\begin{equation}
\sigma_{\nu} = a_{\nu_o} \ H(a,x)
\end{equation}

\noindent
where $H(a,x)$ is the Hjerting function

\begin{equation}
H(a,x) = {a \over \pi} \ \int^{\infty}_{-\infty}{ {exp(-y^2)} 
\over {(x-y)^2 + a^2}} dy 
\end{equation}

\noindent
and 

\begin{equation}
a_{\nu_o} = { {\sqrt{\pi} e^2} \over { m_e c}} { f \over {\Delta \nu_D}}
\end{equation}

\noindent
where we have defined the Doppler frequency, $\Delta \nu_D$, 

\begin{equation}
\Delta \nu_D = { 1 \over {\lambda_o} } \sqrt{ {2 k T} \over {m_{atom}}}
\end{equation}

\noindent
which is the central frequency of the Lorentzian of an atom with the 
RMS velocity of the Maxwellian.  The dimensionless frequency $x$ is

\begin{equation} 
x = { {\nu - \nu_o} \over \Delta \nu_D} = { {\lambda - \lambda_o} \over 
{\Delta \lambda_D}}
\end{equation}

\noindent
the difference between the frequency or wavelength and the line center in
units of the Doppler frequency or Doppler wavelength, 

\begin{equation}
\Delta \lambda_D = { {\lambda_o^2} \over c} \Delta \nu_D = 
{ {\lambda_o} \over c} \sqrt{ {2kT} \over {m_{atom}}}
\end{equation}

The Voigt integral cannot be evaluated analytically and so is generated numerically.
Humlicek (1979) gives a fortran subroutine for $H(a,x)$, and numerical tables are
given by Finn $\&$ Mugglestone (1965).  
In the quasar absorption line literature,
the program {\it VPFIT} is in widespread use for fitting Voigt profiles.  
{\it VPFIT} was developed by Bob Carswell and several generations of students, 
who have
generously made it available to the community (Carswell et al. 1995). 
Theoretical Voigt profiles are shown in Figure 2.  

\begin{figure}
\centerline{\psfig{figure=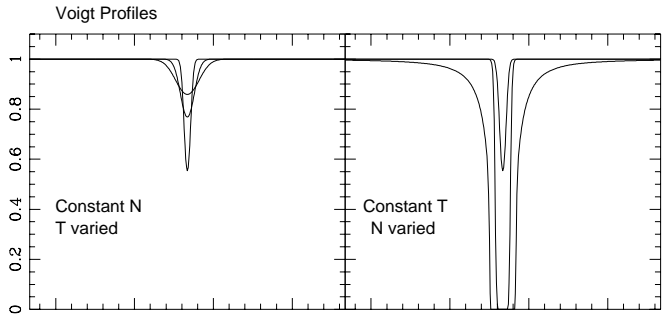,width=10cm,angle=0}}
\caption{Theoretical Voigt profiles.  Left side shows lines on
the linear part of the curve of growth, $\tau_{\nu _o}<1$, 
for constant column density $N$ and $T=10,000$; 50000 and 150000 K.  
Right hand side 
shows profiles for constant $T$ and increasing $N$: $\tau_{\nu _o}=$0.5, 23 
(flat or saturated) and 23,000 (damped).}
\end{figure}

The observed line profile depends on the dimensionless optical depth at 
line center, $\tau_{\nu_0}$, which is

\begin{equation}
\tau_{\nu_o} = \int \ n \ \sigma_{\nu_o} dl
\end{equation}

\noindent
where the right hand side is the integral through the absorbing 
cloud of the volume density $n$ (units are  cm$^{-3}$) times
the cross-section $\sigma_{\nu_0}$ (units are cm$^{2}$) 
at line center for absorption.  
One can show that 

\begin{equation}
\sigma_{\nu_0} = { {\sqrt{\pi} e^2} \over {m_e c} } \ {f \over {\Delta \nu_D}}
\end{equation}
 
\noindent
so if the volume density $n$ is uniform, we can write

\begin{equation}
\tau_{\nu_o} = N { {\sqrt{\pi} e^2} \over {m_e c} } \ {f \over {\Delta \nu_D}}
\end{equation}

\noindent
where N is the {\it column density} with units cm$^{-2}$.  Note that the
optical depth at line center, $\tau_{\nu_o}$, increases with increasing
column density, and is higher for higher $f$-values.  As the temperature
of the atoms increases, $\tau_{\nu_o}$ decreases.  
If $\tau > $ 1 then the line is said to 
be saturated.  Increasing the column density of a saturated line  
has little effect on the line except to make it slightly wider, so
it is difficult to measure the column density when this is the case. 

For most metal-line transitions, and typical interstellar 
conditions, $T\sim10,000$ K,
so that most metal lines have very narrow line widths, and are almost
always unresolved, even at high spectral resolution.  
Because the line width is inversely proportional to the square root of 
the mass of the absorbing atom, hydrogen lines are much wider for the 
same temperature.  

\subsubsection{The curve of growth}

If the spectral resolution of the data is not good enough to resolve the
line profiles, we can still learn a lot by measuring the $``$equivalent 
width" of the absorption line, which is defined as the width of
a rectangular line with area equal to the absorbed area of the actual 
line (and typically is expressed with units of wavelength).  
The equivalent width reduces the information in the line to one
number, proportional to the area of the line, or its strength;  the 
line profile information is lost.  However, if you measure lines 
from the same ion with different $f$-values,
then you can construct a $``$curve-of-growth" and 
deduce the column density, $N$, and temperature, $T$, of the ions in the
absorbing cloud. 

\begin{figure}
\centerline{\psfig{figure=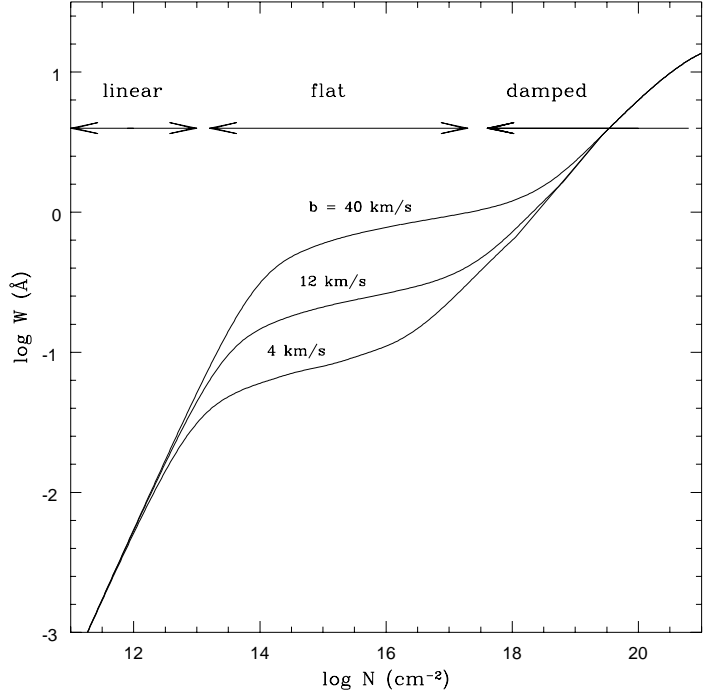,width=10cm,angle=0}}
\caption{Theoretical curve of growth for hydrogen Ly$\alpha$.  Equivalent
width W in \AA\ versus column density N.  Linear,
flat and damping parts of the curve of growth are indicated. }
\end{figure}

A theoretical curve-of-growth is show in Figure 3.  
On the linear
part of the curve of growth, the equivalent width 
increases linearly with increasing column density. On the flat 
part of the curve of growth, the lines are
saturated at line center.  The equivalent width is then 
insensitive to column density but has some sensitivity to
temperature, or equivalently, the $b$-value of the line, where 

\begin{equation}
b = {\sqrt{{{2kT} \over m}}}
\end{equation}

\noindent and has units of km s$^{-1}$.

A convenient rule of thumb is that
for a line on the linear part of the curve of growth  

\begin{equation}
log({{W_{\lambda}} \over {\lambda}}) = log({{N}\over{N_H}}) + log \lambda f + 
log N_H - 20.053
\end{equation}
 
\noindent
where W$_{\lambda}$ is the equivalent width, ${N}\over{N_H}$ is the abundance
of the atom relative to hydrogen, N$_H$ is the total hydrogen column 
density and $\lambda$ is the wavelength of the transition in Angstroms. 

From equation 2.13  we see that ultraviolet absorption lines are very 
sensitive to small amounts of gas.  For example, C III is the dominant
ion for many interstellar 
clouds, and has a strong line at $\lambda$=977 \AA, with
$f=0.768$, so that log $\lambda$ f = 2.876.  The abundance, log C/H = -3.35 for
solar abundance gas.  One can easily detect a line with W$_{\lambda}=0.1 \AA$,
which corresponds to a log $N_H$ = 16.5 cm$^{-2}$.  This is a very small column
compared to that easily probed by 21-cm or other emission line techniques. 
For hydrogen Ly$\alpha$, column densities of log $N_H$ = 12.5 cm$^{-2}$ are
easily measured.

\subsubsection{Doublet Ratios}

If a full curve-of-growth cannot be constructed, it is often sufficient
to measure the {\it doublet ratio} of so-called H and K lines of a particular 
ion. These are pairs of lines which  
are near each other in wavelength so are easy to observe simultaneously,
but far enough apart that they are easily resolved, even at moderate spectral 
resolution.  Because they unambiguously identify redshift systems with
two lines, doublets have been exploited to survey metal absorbers -- 
particularly the very strong 
Mg II $\lambda\lambda$ 2803, 2796 \AA\ and C IV $\lambda\lambda$ 
1548, 1550 \AA\ transitions.  Other doublets include 
Si IV $\lambda\lambda$ 1393, 1402 \AA\, N V $\lambda\lambda$ 1238, 1242\AA\,
O VI $\lambda\lambda$1031, 1037 \AA.  In the optical, 
Ca II H\&K $\lambda\lambda$3933, 3968 \AA\ 
and Na I D $\lambda\lambda$5889,5895 \AA\ are doublets seen 
in absorption. 
 
The shorter wavelength member (H) of the doublet has an $f$-value 
which is twice
the $f$-value of the longer wavelength member (K) of the doublet.  Thus, if
both lines are on the linear part of the curve of growth, the ratio of
the equivalent widths of the two lines will be exactly the ratio of 
the $f$-values, or 2.   In this case,
one can derive the column density from the equivalent width of 
either line, but not the $b$-value.  At higher columns,
the K line can be on the flat part of the curve of growth, with the H line
is on the linear part.  Then the ratio of the equivalent widths will be
between 1 and 2.  The column density may be derived from the equivalent 
width of the H line,
and the $b$-value from the equivalent 
width of the K line.  At still higher columns, both H and K 
are saturated.  The equivalent widths of both lines are equal and the column
density is not possible to derive (although one can measure a lower limit  
assuming a $b$-value).  The doublet ratio and curve-of-growth are useful
checks even if the spectra have sufficient resolution to fit Voigt profiles. 

The wavelengths of doublets in redshifted quasar
spectra can be used 
to investigate whether the fine structure constant, $\alpha$,
changes with cosmic epoch (Savedoff 1956; 
Bahcall, Sargent \& Schmidt 1967; 
Wolfe, Brown \& Roberts 1976;
Ivanchik, Potekhin \& Varshalovich 1999;
Levshakov 1994; Cowie \& Songaila 1995; 
Webb et al. 2000;
Dzuba, Flambaum \& Webb 2001,
Murphy et al. 2000abc and references therein).
Murphy et al. (2000a) at al find a change significant at the 
4$\sigma$ level, with 
$\alpha$ being smaller in the past.  The implications for physics 
are described in Murphy et al. (2000b).   
Various sources of systematic error may be important
for accessing the security of this result (Murphy et al. 2000b). 
Radio observations of 21cm hyperfine splitting in redshifted 
absorption also put limits on
the evolution of $\alpha$ (Drinkwatter et al. 1998;
Carilli et al. 2000 and references therein). 

\subsubsection{Limitations}

Although ultraviolet absorption lines are remarkably sensitive probes
of high redshift gas, there are several limitations to absorption line
studies worth keeping in mind.  Heavily reddened lines  
of sight are impossible to observe in the rest frame UV, and most 
absorption line studies have used quasars selected to be bright and blue.
Note that half the mass of the interstellar medium 
in the Milky Way at the solar circle
is in molecular clouds; these are completely inaccessible to ultraviolet
absorption studies.  Thus we expect quasar absorbers to select
against reddened lines of sight (Malhotra 1997 and references therein).
Likewise, very hot gas is also selected against, since the atoms 
are all ionized and so there are no accessible absorption lines.  

Although
many absorption features are well fit by a Voigt profile with some column 
density $N$ and temperature $T$, the absorption is really a pencil beam 
average of the density and temperature along the line of sight. 
The volume density may be estimated by measuring fine-structure 
lines from excited states such as Si II $^{*}$  
or C II$^{*}$ and comparing  them to the columns of the ground state
(Bahcall \& Wolf 1968; Sarazin et al. 1979).
If the excitation is caused by collisions, the ratio of the excited
state columns to ground state columns will be sensitive to density. 
However, these lines are usually heavily saturated. Also, the excited
states may be populated by direct excitation by infrared photons, 
or indirect excitation by ultraviolet photons into excited states which
subsequently decay.  Thus the interpretation of the observations 
requires some knowledge of the radiation field.  In fact, usually 
the excited fine structure lines in quasar 
absorbers are used to estimate the radiation field, not the density 
(see below). 

There is a small literature on 
the validity of curve of growth techniques and
Voigt profile fitting generally, and their application to 
quasar absorbers in particular.   At issue is the ability 
to properly account for multiple velocity components, particularly for
saturated lines.  The perils of applying the curve of growth to 
blended features are described by Nachman \& Hobbs (1973), Crutcher (1975),
Gomez-Gonzalez \& Lequeux (1975), Parnell \& Carswell 1998). 
A more optimistic view is described by Jenkins (1986).   Levshakov and 
collaborators have written a series of papers emphasizing the
effect of turbulent velocities on the interpretation of absorption line
data (Levshakov \& Kegel 1997; Levshakov et al. 1999 and references therein). 
As with all astronomical measurements, the limitations of the observations
need to be understood and dealt with as best as possible.
 
\begin{figure}[htb]
\centerline{\psfig{figure=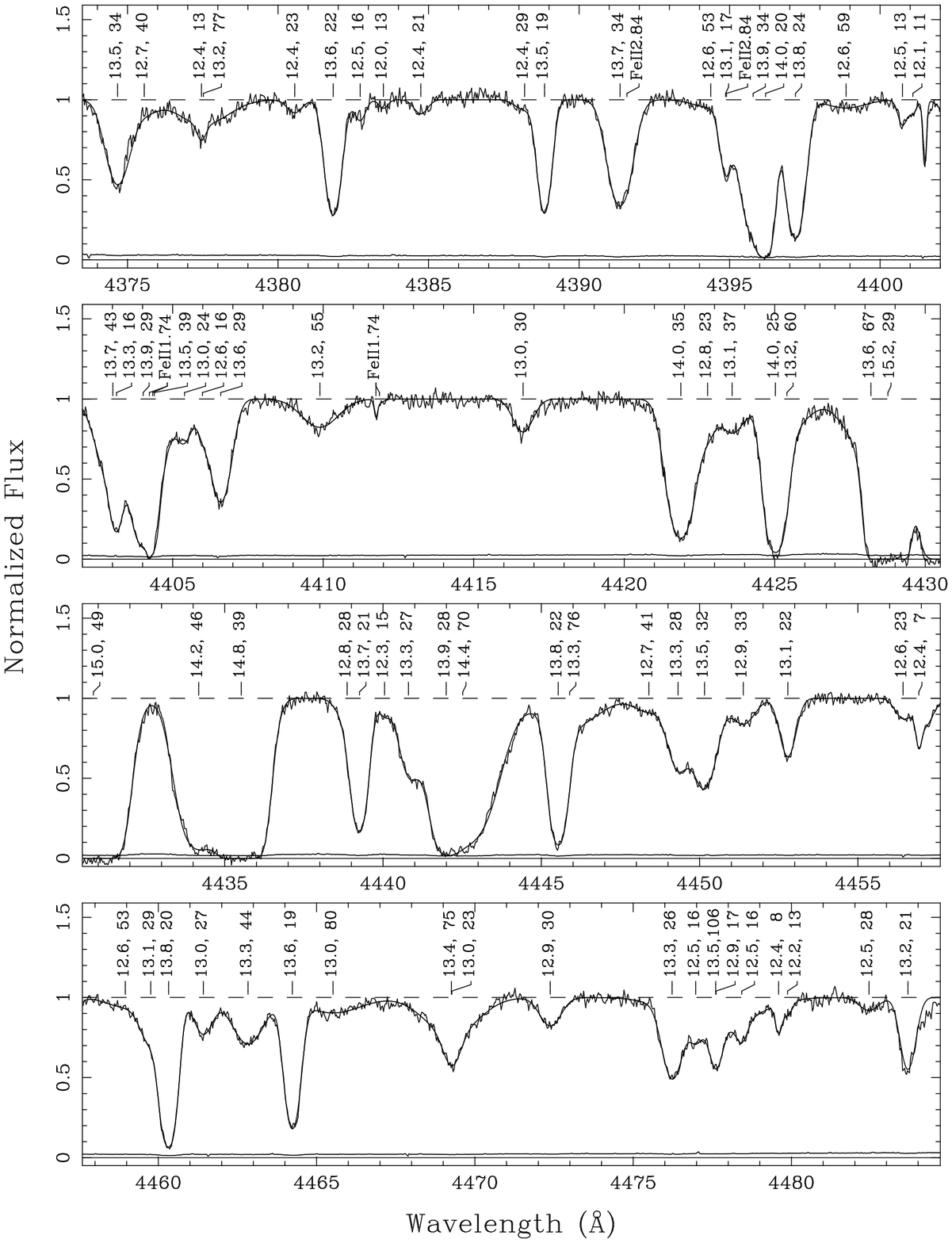,width=10cm,angle=0}}
\caption{Voigt profile fits to a quasar spectrum. The spectrum is normalized
to a continuum level of one.  Wavelengths are vacuum
heliocentric values.  The log column density (in log
cm$^{-2}$) and velocity dispersion (in km s$^{-1}$) of each line are
displayed above a tick mark indicating the center of the line.
Identified metal ions are indicated along with the redshift of
the metal system.  The calculated 1$\sigma$ error spectrum
(per pixel) is shown just above zero on the same scale.  The
pixel size is 2 km s$^{-1}$.  
From Kirkman \& Tytler (1997).}
\end{figure}

Voight profiles fit observed absorption line spectra quite well (Figure 4); 
however, they may be regarded as just one
convenient parameterization of the data.  The traditional 
approach has been to construct 
a list of lines, measure their redshifts, equivalent widths, column 
densities and b-values, and then base
subsequent analysis on the line list.  Alternatively, one can compare 
optical depths of the different transitions of the absorbing line profile
on a pixel-by-pixel basis (Savage \& Sembach, 1991).
 In the case of the intergalactic medium, 
it has been productive to mitigate the effects of blending by comparing
 the observed spectra on 
a pixel-to-pixel basis with the simulated $``$observations" of 
theoretical models for the intergalactic medium (Press \& Rybicki 1993;
Press, Rybicki \& Schneider 1993;
Dobrzycki \& Bechtold 1996; Weinberg et al. 1999).  
Outram, Carswell \& Theuns (2000) argue that Voigt profiles are a poor representation
of Ly$\alpha$ forest lines and do not represent the proper physical
picture.  We discuss the Ly$\alpha$ forest further below.

\subsection{Echelle Spectrographs}

In order to obtain high resolution spectra (spectral resolution $R = \lambda / 
\Delta \lambda \sim$ 30,000 or greater) astronomers use echelle spectrographs,
where the dispersing element is a grating used in high spectral order
and angle of diffraction.  A second prism or grating serves as a cross-dispersing
element which separates the closely spaced spectral orders perpendicular
to the dispersion direction of the echelle grating.
A good description of astronomical spectrographs 
is given in Schroeder's textbook {\it Astronomical Optics} (2000), 
and a review of
the echelles planned or in use at large telescopes can be found in Pilachowski 
et al. (1995).  Most of the results reviewed here were obtained with
Vogt's HIRES spectrograph on the Keck 10m telescope (Vogt et al. 1994).

\section{Absorption by Material Associated with the QSO}

About 10\% of QSOs show absorption attributable to gas 
associated with the QSO central engine or its immediate environs.   
In fact, the first quasar absorber reported 
turned out to be an $``$associated" C IV absorber (Sandage 1965).  
This category includes the radiatively driven winds producing the 
broad absorption line (BAL) QSOs 
(Turnshek 1995 and references therein, see also Becker et al. 2000; 
Green et al. 2000), 
the associated C IV absorbers (Foltz et al. 1986, Anderson et al. 1987), 
the associated Mg II absorbers (Low, Cutri, Kleinman \& Huchra 1989; 
Aldcroft, Bechtold \& Elvis 1994), and the X-ray warm (that is, ionized) 
absorbers (Fiore et al. 1993, Mathur et al. 1994, George et al. 1998).  
An attempt to understand the various 
observations in terms of the unified model for the central engine 
is given by Elvis (2001). 
Associated absorption in radio galaxies is 
discussed in these proceedings by Rawlings. We have not reviewed this
subject here, but the interested reader can start with Arav, Shlosman 
\& Weymann 1997. 

\section{The Lyman Alpha Forest}

The techniques described above have been applied to large
numbers of spectra of high redshift quasars in order to study the
weak Ly$\alpha$ lines -- those with log N(HI) $\sim$ 12.5-16.5 cm$^{-2}$.
Prior to about 1995, the Ly$\alpha$ forest was defined as 
$``$Ly$\alpha$ lines with
no metal absorption at the same redshift".  However, it had long been 
suspected that the quasar absorbers discovered by their Ly$\alpha$
absorption should not be divided rigidly between $``$metal absorbers" 
and $``$forest" -- that is, absorbers associated with galaxies and absorbers  
associated with intergalactic gas (e.g. Tytler 1987). 
The data obtained with 4m class telescopes would not have detected 
metal line absorption for the low column Ly$\alpha$ clouds even
if they had solar metallicities -- the strong lines of C III
O VI, C IV or Si IV were just too weak
(Chaffee et al. 1986).   When C IV and Si IV metal absorption was 
detected for the stronger Ly$\alpha$ forest lines (Cowie, Songaila, 
Kim \& Hu 1995;
Womble, Sargent \& Lyongs 1996), the distinction
between $``$Ly$\alpha$ forest" and $``$metal absorber" became 
less clear.   However, as described below, 
the low column Ly$\alpha$ absorbers differ in
several key observed properties from those at the opposite 
end of the column density distribution, the damped Ly$\alpha$ absorbers.
The distinction between forest and metals may be a function
of redshift as well.    
 
\begin{figure}
\caption{Simulated Ly$\alpha$ forest at $z=3$ showing non-spherical
geometry for the absorbing gas.  Three-dimensional isodensity surfaces for 
baryon density three times the local mean.  From Cen \& Simcoe (1997).}
\end{figure}

Around the same time, numerical codes by several groups produced the
first simulations of
the evolution of the intergalactic medium and gas in galaxies 
at high redshift (Cen, Miralda-Escude, Ostriker \& Rauch 1994; 
Zhang, Anninos \& Norman 1995;
Hernquist et al. 1996;
Wadsley \& Bond 1996;
Theuns et al. 1998).  These allowed a detailed examination of the structure
and dynamics of quasar absorbers, and produced a visual picture of what the
structures are: typically sheets and filaments, aligned with
the sites of future galaxies (see e.g. Figure 5).   
Numerical techniques used to model the evolution of galaxies and large scale
structure were described in detail by Simon White at the Winter School.
Here we focus on the observations, and the theoretical results 
which have shaped our understanding of the quasar absorbers.

\subsection{The Column Density Distribution}

One basic observed quantity is the distribution of
column densities in the Ly$\alpha$ clouds.  Voigt fits to
high dispersion spectra have been used to derive the distribution
from log $N(HI)$ = 12.3-14.5 cm$^{-2}$ (Hu et al. 1995; Lu, 
Sargent, Womble \& Takada-Hidai 1996;
Kim, Hu, Cowie \& Songaila 1997; Kirkman $\&$ Tytler 1997).
The distribution is a power-law, written 

\begin{equation}
f(N_{HI}) = 4.9\times 10^7 \ {N_{HI}}^{-1.46}
\end{equation}

\noindent
where $f(N_{HI})$ is 
defined as the number of absorbing systems per unit redshift path 
per unit column density as a function of neutral hydrogen column density, 
$N_{HI}$.  The redshift path $X(z)$ is

\begin{equation}
X(z)= { 2 \over 3} [ (1+z)^{3/2} -1 ]
\end{equation} for $q_o = 0.5$.
The data are shown in Figure 6.
 
\begin{figure}[htb]
\centerline{\psfig{figure=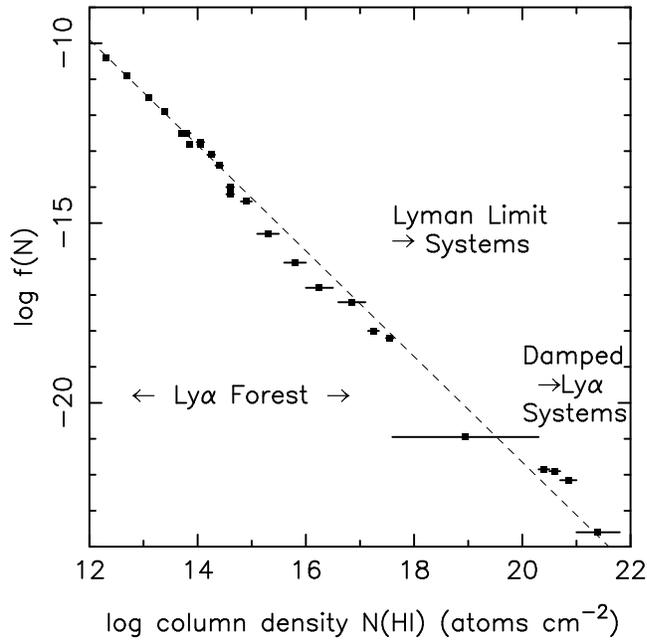,width=10cm,angle=0}}
\caption{ The column density distribution function of neutral
hydrogen for 12 $\le$ log N(HI) $\le$ 22, for the Ly$\alpha$ 
forest, Lyman limit systems and the damped Ly$\alpha$ absorbers.
The dashed line is a power law, $f(N) \propto N^{-1.46}$,
which fits the data reasonably well, 
over 10 orders of magnitude in column density.
From Storrie-Lombardi \& Wolfe (2000).}
\end{figure}

\noindent
Remarkably, the extension of this power law connects pretty well through
the metal absorbers down to the damped Ly$\alpha$ absorbers, with log $N(HI) = 22.0$ cm$^{-2}$ (Tytler 1987, Petitjean et al. 1993; Storrie-Lombardi
\& Wolfe 2001).

\subsection{The b-value distribution}

The distribution of Doppler $b$-values are roughly Gausians 
with a mean of about 30 km s$^{-1}$ and width 8 km s$^{-1}$ 
(Hu et al. 1995; Lu et al. 1996, Kim et al. 1997; figure 7).  The distributions
must be corrected slightly for blending via analysis of  
simulated spectra.  There is also a cut-off b-value, $b_c$, 
on the low side, indicating that below 15 $km s^{-1}$ or so, no narrower Voigt 
components are needed to fit the observations.  

\begin{figure}[htb]
\centerline{\psfig{figure=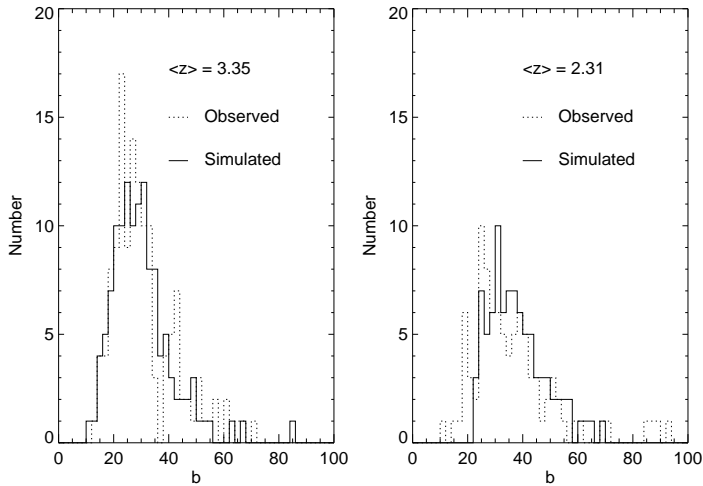,width=10cm,angle=0}}
\caption{The $b$-value distribution of the Ly$\alpha$ forest.
Observed (dotted) and simulated (solid) histograms of the 
$b$-values derived from Voigt profile fitting of Ly$\alpha$ forest
clouds.
From Kim et al. (1997).}
\end{figure}

There is some indication that the mean $b$ and cut-off $b$ decrease with 
increasing redshift (Williger et al. 1994; Lu et al. 1996; 
Kim et al. 1997; Schaye et al. 2000).  On the face of it, this would
suggest that the temperatures and/or the turbulent
broadening of the absorbing gas is increasing with decreasing redshift.    

However, there is more to the story (Schaye et al. 1999, 2000;  
Ricotti, Gnedin \& Shull, 2000). On small spatial scales, the temperature of
the gas is determined by a balance between photoionization heating
by the UV background, and adiabatic cooling from the expansion
of the Universe.  As the the Universe expands, the 
overdensity corresponding to a fixed observed column density decreases
with redshift  (Schaye et al. 1999, 2000; 
Ricotti, Gnedin \& Shull 2000).  The temperatures and densities
obey what these authors call an $``$equation of state",

\begin{equation}
T = T_o ({{\rho} \over {\bar{\rho}} })^{\gamma - 1}.
\end{equation}  

If the IGM is reionized by sources of UV radiation quickly with respect
to the timescales for cosmic 
expansion, the gas is isothermal and $\gamma \sim 1$.
Subsequently, the mean temperature $T_0$ decreases as the Universe
expands, and the slope $\gamma$ increases because higher density
regions expand less rapidly than average regions and photoionization heating
is more effective.  The data show that the gas was nearly isothermal
at $z\sim 3$, corresponding to the epoch of He II reionization (seen
in the observations, section 4.8 below).

\subsection{Evolution with Redshift}

The number of 
lines per redshift observed, dN/dz, can  be written

\begin{equation}
{ {dN} \over {dz} } = { {c n_o(z) \sigma(z)} \over {H_o}} 
{ {(1+z)} \over { (1+q_o z)^{1/2}} }
\end{equation}

\noindent
where

$n_o(z)$ = the comoving number density of absorbers, and

$\sigma(z)$ = the geometric cross-section for absorbtion.

\noindent
Note that

\begin{equation}
 n_o(z) \sigma(z) = { 1 \over l}
\end{equation}

\noindent
where $l$ = the mean free path for absorption.

For objects with no intrinsic evolution of $n_o(z) \sigma(z)$,
and $\Lambda$=0, 
we have 

\begin{equation}
{ {dN} \over {dz} } \propto 1+z 
\end{equation}

\noindent for $q_0=0$ and

\begin{equation}
{ {dN} \over {dz} } \propto (1+z)^{1/2}
\end{equation}

\noindent for $q_0=0.5$.  The observations can be fit with a function

\begin{equation}
{ {dN} \over {dz} } = N_o (1+z)^{\gamma}
\end{equation}
 
The results are shown in Figure 8.
The extension of $dN/dz$ from $z=2$, observable from the ground,
to $z=0$ was one of the long-anticipated results of the $HST$ 
quasar absorption line Key Project (Bahcall et al. 1993; Schneider et al. 1993;
Savage et al. 1993; Bergeron et al. 1994;
Stengler-Larrea et al. 1995;
Bahcall et al. 1996;
Jannuzi et al. 1998;
Wemann et al. 1998;
Savage et al. 2000).  
Owing to the small numbers of lines in early ground-based
samples, it has become customary to fit the observed dN/dz without binning
into redshift bins. The best fit $\gamma$ for lines stronger
than some threshold equivalent width is solved for using maximum likelihood  
techniques (Murdoch, Hunstead, Pettini \& Blades 1986; Weymann et al. 1998).  

\begin{figure}
\centerline{\psfig{figure=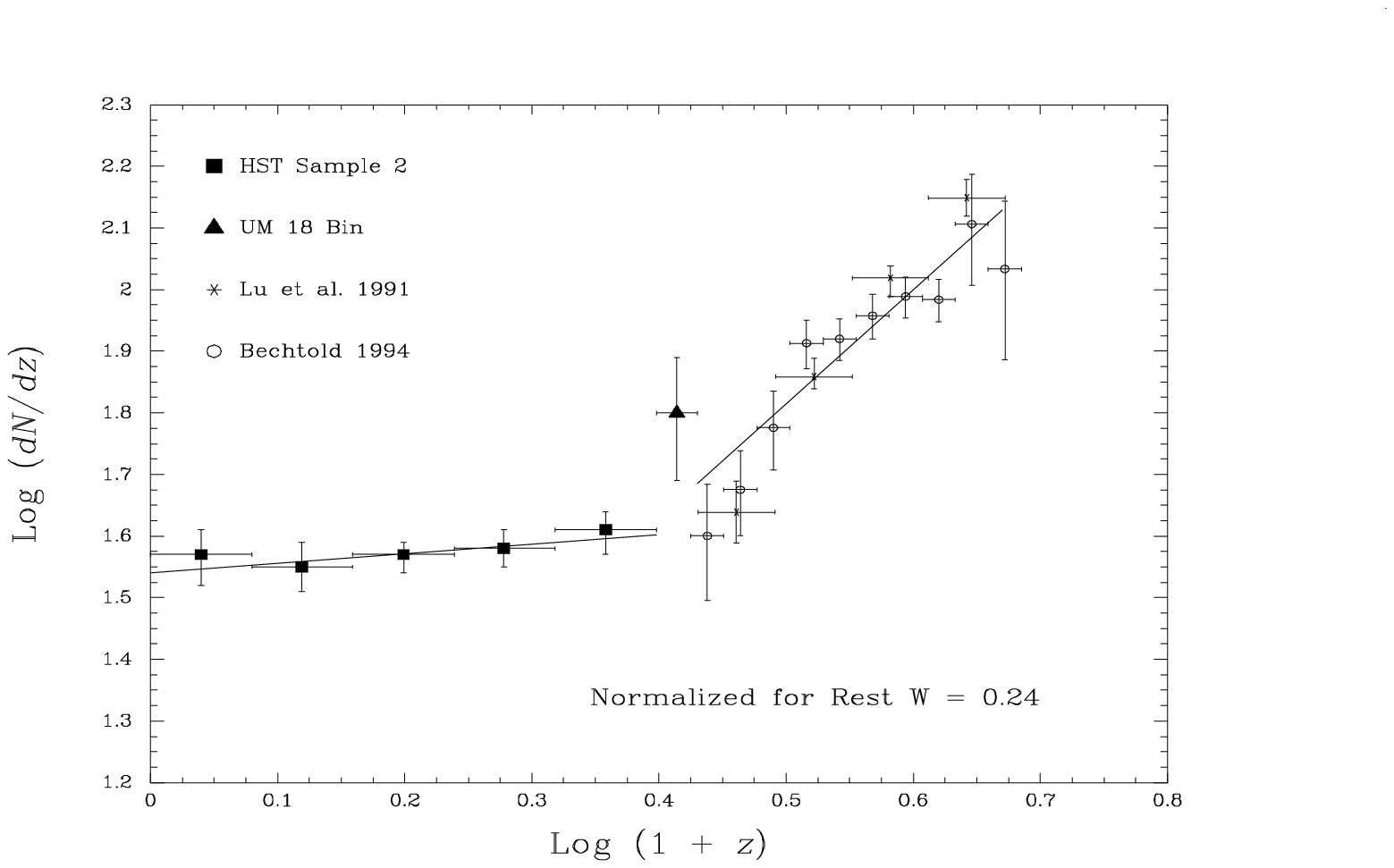,width=12cm,angle=0}}
\caption{Evolution of the Ly$\alpha$ forest.
At $z<1.6$ the data come from
the HST Key Project catalogue of Ly$\alpha$ absorbers, 
Bahcall et~al.~1996, Jannuzi et al. 1998.
At higher redshifts two ground-based samples 
are shown (Lu, Wolfe \& Turnshek 1991; Bechtold 1994). 
Results from high dispersion data (Kim et al. 1997) are similar.
The separate fits to the HST and ground-based data are 
shown as the two solid lines, with slopes of
$\gamma=0.5$ and $1.85$ respectively.
From Weymann et al. (1998).}
\end{figure}

Figure 8 shows that for $z>2$, $\gamma \sim 2-3$, that is the evolution
is very steep, in the sense that there was more absorption in 
the past.  There is some indication that $\gamma$ steepens for $z>4$ 
(Williger et al. 1994).  At $z=1-2$ the distribution flattens,
and at $z<1.5$, $dN/dz$ is consistent with $``$no evolution".  

Qualitatively, the evolution can be understood in terms of 
three main factors. 
One expects that the collapse of structures and the mergers of 
protogalactic fragments decreases both $n_o$ and $\sigma$, so that
$dN/dz$ decreases with $z$.  Second, the 
the UV radiation field is expected to decrease from $z\sim2$ to $z=0$, 
whether it is comprised of photons from quasars or star-forming galaxies.
Thus as the UV field decreases the neutral fraction increases and the 
number of detected lines increases.  This tendency is balanced by  
the third effect, which is that as cosmic expansion proceeds, the density of
the gas decreases and even if the photoionizing UV field were constant,
the neutral fraction would decrease and the number of Lyman alpha lines
detected will increase.  Undoubtedly the evolution at low $z$ is even
more complicated than this. 

\subsection{The UV Radiation Field and the Proximity Effect}

A key quantity for understanding the Ly$\alpha$ forest is
the ionizing ultraviolet radiation field.  Arons \& McCray (1969) 
and Rees and Setti (1970) suggested that quasars are
likely to be a significant source of hard UV photons at high redshifts.
Star forming galaxies may also contribute (Bechtold et al. 1987; Haardt
\& Madau 1996; Steidel, Pettini \& Adelberger 2001).
The number counts of Ly$\alpha$ forest clouds themselves provide
an independent way to estimate the UV background by analysis of the
so-called proximity effect.  

Weymann, Carswell and Smith (1981) first pointed out that
the number density of Ly$\alpha$ forest lines
is systematically low near the quasar being used to probe them.
They suggested that photoionization
by the quasar's UV light might be responsible. 
Carswell et al. (1987) pointed out that this 
 effect could be used to estimate the ambient UV background.
Qualitatively, the extent of the proximity effect indicates roughly 
how far the ionizing radiation from the quasar is exceeds that of the 
background.  By measuring the brightness of the quasar (often 
the Lyman limit flux is directly observable), one can therefore estimate
the value of the specific intensity of the background, J$_{\nu}$.  

Initially, the estimates of J$_{\nu}$ derived from proximity effect
analysis of various high redshift samples indicated high 
values.
For example, Bajtlik, Duncan \& Ostriker (1988) 
found  that log J$_{\nu}$ = -21.0 $\pm$ 0.5
ergs sec$^{-1}$ cm$^{-2}$ Hz$^{-1}$ sr$^{-1}$
for the redshift range of their sample, $1.7 < z < 3.8$.   
This was significantly larger than estimates of the quasar background, 
particularly at $z>$3 (e.g. Bechtold et al. 1987). 
Bajtlik, Duncan \& Ostriker (1988) suggested that there was another
source of ionizing radiation, young hot stars in galaxies 
(Bechtold et al. 1987,
Miralda-Escude and Ostriker 1990) or a population of quasars which we don't 
see today because they are obscured by dust in intervening galaxies 
(Ostriker and Heisler 1984; Wright 1986; Heisler and Ostriker 1988ab; 
Boisse and Bergeron 1988; Najita, Silk and Wachter 1990; Wright 1990; 
Ostriker, Vogeley and York 1990; Fall, Pei and McMahon 1989, 
Pei, Fall \& Bechtold 1991; Pei \& Fall 1995, Pei, Fall \& Hauser 1999).  

Subsequently, many quasars with z$>$3 were found, and the luminosity function
for high redshift quasars became better defined. 
The value for the quasar contribution to the 
background increased
(Miralde-Escude and Ostriker 1990; Haardt \& Madau 1992),
and is now much closer to the proximity effect estimates.
Also, the sources of various systematic effects in proximity effect 
estimates of J$_{\nu}$ 
have been dealt with (Bechtold 1995; Scott, Bechtold \& Dobrzycki 2000a).
The proximity effect estimates of $J_{\nu}$ are now in broad agreement
with the predictions of the contribution of known quasars to the 
ionizing background (Bechtold 1994, Lu et al. 1991; Scott et al. 2000b).
 
\begin{figure}[htb]
\centerline{\psfig{figure=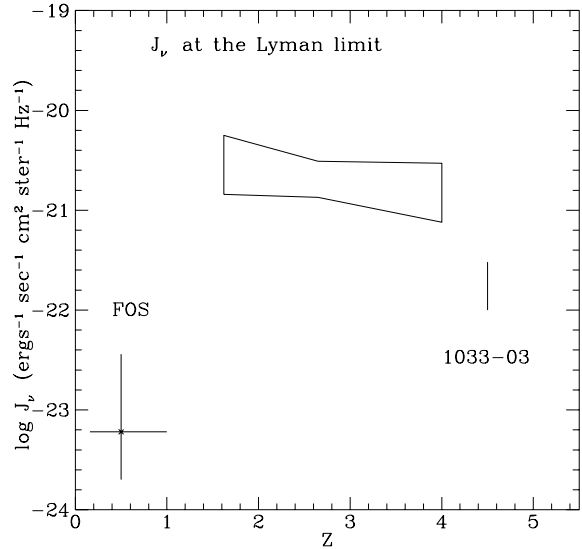,width=8cm,angle=0}}
\caption{Evolution of the ionizing UV radiation field. J$_{\nu}$, 
as derived from the proximity effect, as a 
function of redshift, $z$.  From  
Bechtold (1995).}
\end{figure}

At $z<2$, one expects $J_{\nu}$ to decline because of the decline
in the population of quasars and star-bursting galaxies. 
Kulkarni \& Fall (1993) derived a 
value based on early HST Key Project samples, and a larger sample has
been analyzed recently by Scott et al. (2001).  
A summary of $J_{\nu}$ at the Lyman limit as derived from
the proximity effect is shown in Figure 9. 
While the uncertainties are still large, broadly speaking, the results
are consistent with the predicted contribution from quasars.
The general trend that the ionizing UV radiation field is smaller at z$\sim$0
and z$\sim$4.5 than at z$\sim$3 is probably secure.

If quasars happen to lie along the line-of-sight they  
could cause observable voids in the Ly$\alpha$ forest far from 
the quasar (Paczynski 1987, 
in Bajtlik, Duncan \& Ostriker 1988). 
Although luminous quasars are 
rare objects, the path 
length sampled by a typical quasar spectrum is long, and such chance 
coincidences are common enough to be interesting.
Kovner and Rees (1989) discussed possible $``$clearings" 
in the Ly$\alpha$ forest from the $``$two quasar proximity effect" and 
concluded that the unsuccessful searches for voids in the Ly$\alpha$
forest up to that time were consistent with expectations (Carswell and 
Rees 1987;  Ostriker, Bajtlik \& Duncan 1988; 
Duncan, Ostriker and Bajtlik 1989).  In principle, one could 
use the $``$two quasar proximity effect" to estimate
quasar lifetimes since a population of numerous short-lived ionizing 
sources would cause fewer large voids than rare, long-lived ones.  
Also, if the UV light from quasars is strongly 
beamed, one would expect to see examples of bright quasars along the line
of sight with no associated proximity-induced void, or voids in the Ly$\alpha$ 
forest with no observable quasar to cause them.  Of course voids could also 
be present from real underdensities in the total hydrogen gas, and these 
would be difficult to distinguish from a proximity-induced void.
Dobrzycki and Bechtold (1991ab) found one significant void in the 
Ly$\alpha$ forest.  There is a known quasar nearly 
but not {\it quite} at the right redshift, close enough to the 
line-of-sight, with the right brightness, to cause it.  
Subsequent searches for a more suitable ionizing quasar have been fruitless.  
Fernandez-Soto et al. (1995) looked for  proximity-induced
voids in the spectra of 3 quasars with known quasars along the line-of-sight, 
but with only 3 objects, they were not able to tell one way or the other 
whether significant voids are present.  

Zuo (1992ab) generalized these ideas and looked for underdense regions 
in the forest, rather than complete voids.  He pointed out that in principle, the nature of the 
sources of the ionizing background could be deduced from the statistics 
of underdense regions.  A fixed {\it integrated} background  could be 
made up of rare, bright sources (e.g. luminous quasars) or numerous, 
fainter ones (e.g. star-forming galaxies): these would cause different 
signatures in the distribution of Ly$\alpha$ clouds through the proximity 
effect.  Unfortunately, the data available did not allow strong 
statements to be made.

\subsection{Characteristic Size and Geometry}

The characteristic size and geometry of the absorbing
structures can be estimated by searching for common absorption
in the spectra of pairs or groups of quasars which are serendipitously
near each other on the sky.  Individual images of gravitationally
lensed quasars can also be used, provided the lensing geometry is known, 
since each image traverses a different
path through space.  

Lenses probe very small separations, and 
generally show that all the lines seen in one image are seen in
the other, with very strongly correlated equivalent widths 
(Young et al. 1981; 
Weymann \& Foltz 1983; 
Foltz, Weymann, Roser \& Chaffee 1984; 
Smette et al. 1992, 1995;  
Turnshek \& Bohlin 1993;
Bechtold \& Yee 1995;
Zuo et al. 1997; 
Michalitsianos et al. 1997; 
Petry, Impey \& Foltz 1998; 
Lopez et al. 1999;
Rauch, Sargent \& Barlow 1999).  Thus the characteristic sizes must
be larger than the typical separations.  For very wide pairs of 
quasars, where the projected separation is 20$"$ to arcminutes,
few if any lines are seen in common (Sargent et al. 1980; 
Crotts \& Fang 1998).  Thus the sizes
are smaller than the separations probed by large pairs.  Pairs and groups
of quasars with separations of 10$"$ to 1 arcminute contain lines in
common, and lines which are not in common, 
and so are probing the characteristic sizes of the 
absorbing structures 
(Bechtold et al. 1994; 
Dinshaw et al. 1994, 1995;
Fang et al. 1996;
Monier, Turnshek \& Lupie 1998;
Monier, Turnshek \& Hazard 1999).     

Unfortunately, the number of favorable groupings of quasars bright enough
to observe is extremely
small, since luminous quasars are themselves quite rare objects. 
One of the useful products of the Sloan digital Sky Survey will be
the discovery of groups and pairs of quasars to observe for common absorption. 
 Nevertheless, it is
clear from existing observations that the 
Ly$\alpha$ clouds have sizes 200-500 $h_{100}^{-1}$ kpc, which are very large.

Note that common absorption measures essentially {\it transverse}
sizes.  Rauch \& Haehnelt (1995) pointed out that if the clouds are
spherical and ionized as expected by the UV radiation field of quasars 
(so that the measured N(HI) is a small fraction of the total H),  
then the mass density of the forest would exceed the baryon density 
derived from primordial  nucleosynthesis.  Therefore, the geometry
is more likely sheetlike, as suggested already by numerical simulations.

\subsection{Clustering of the Ly$\alpha$ forest clouds}

One of the early results which distinguished the Ly$\alpha$ forest
clouds from higher column density clouds was the apparent weak 
clustering in their redshift distribution (Sargent et al. 1980; 
Rauch et al. 1992;
Hu et al. 1995;
Cristiani et al. 1995;
Lu et al. 1996;
Fernandez-Soto et al. 1996;
Kirkman \& Tytler 1997).  

\begin{figure}[htb]
\centerline{\psfig{figure=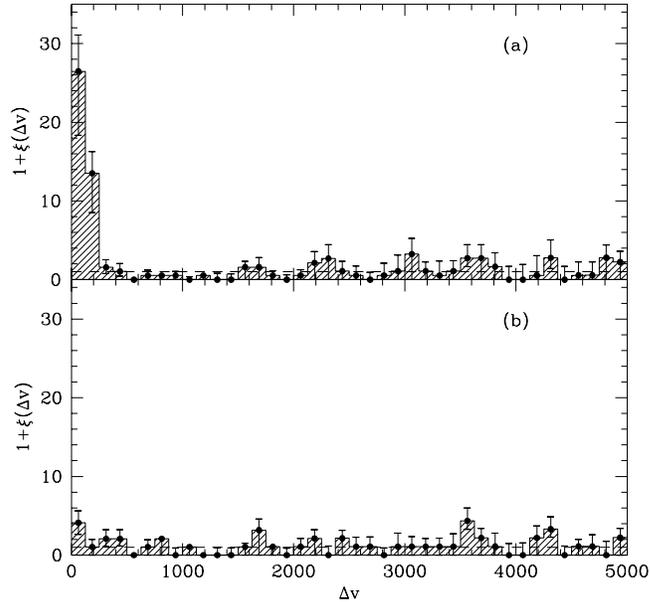,width=10cm,angle=0}}
\caption{Clustering of absorbers at $z\sim 2.5$.  
Two-point correlation function 
as a function of velocity splitting, $\Delta v$, for C IV absorbers 
(a), and Ly$\alpha$ absorbers (b).  From Fernandez-Soto et al. (1996).}
\end{figure}

However, the b-values of the lines are $\sim$25 km s$^{-1}$,
compared to the expected
velocity dispersion of collapsed protogalaxies of perhaps a few
hundred km s$^{-1}$.  Thus, any signal in the two-point correlation 
function may be weakened by blending.  For the metals with b-values
of only a few km s$^{-1}$ it is easier to resolve separate components with 
small velocity splittings, so signal in the two point correlation function is
easier to detect.  Results for C IV absorbers and  
Ly$\alpha$ absorbers are shown in Figure 10.  In either case, the 
clustering amplitude is small compared to that of local galaxies.  

At low redshift, the $HST FOS$ data set has been analyzed by Ulmer (1996)
who claims that there is strong clustering of the forest lines, but this claim
has been questioned by Dobrzycki et al. (2001) who found weaker clustering 
in a larger sample.
The Key Project made the 
interesting claim that Ly$\alpha$ forest lines were correlated with
metal absorbers, but this was based on a very small number of objects,
and statistical analysis of a larger sample again failed to confirm
this result.

\subsection{The Low Redshift Ly$\alpha$ Forest}

One of the most anticipated observations to be made with
HST was the search for Ly$\alpha$ forest clouds at low redshift, 
so one could carry out a detailed study of the relationship between
the clouds and galaxies.  A number of authors
carried out deep galaxy redshift surveys in the directions
of quasars with high quality HST FOS and GHRS spectra, both optically 
(Morris et al. 1993;
Stocke et al. 1995;
Lanzetta et al. 1995;
Shull, Stocke \& Penton 1996;
Chen et al. 1998;
Tripp, Lu \& Savage 1998;
Ortiz-Gil et al. 1999;
Penton, Stocke \& Shull 2000;
Penton, Shull \& Stocke 2000; Figure 11)
and with 21-cm techniques (Van Gorkom et al. 1996).
 
\begin{figure}[htb]
\caption{Pie-diagram distributions of recession velocity and right ascension  
of bright (CfA survey) galaxies and four Ly$\alpha$ absorbers 
toward Mrk~501 and Mrk~421. Two of these systems lie in voids; 
the nearest bright galaxies lie $>4 h_{75}^{-1}$ Mpc from the 
absorber.  From Shull, Penton \& Stocke (1999).}
\end{figure}

The results show that the forest clouds at low redshift arise in a variety of 
places.  The clouds show some correlation with galaxies but
not aren't as strongly correlated as galaxies are to each other (Stocke 
et al. 1995).  There are some clouds at redshifts which place them
in voids of the galaxy distribution.   Others may be associated with
tidal debris of galaxies or groups of galaxies. 
Although Lanzetta et al. (1995) have argued that the Ly$\alpha$ clouds
arise in huge halos of individual galaxies, a concensus 
is growing that they are associated more often with structures 
in the galaxy distribution, such as superclusters and sheets.

\subsection{The Gunn-Peterson Effect; Metals and the Ionization of the IGM}

Gunn \& Peterson (1965) pointed out that any generally distributed 
neutral hydrogen would produce a broad depression in the spectrum of 
high redshift quasars at wavelengths shortward of Ly$\alpha$ $\lambda$1215.
Since such a depression is not seen, they concluded that 
the intergalactic medium must be ionized -- most of the hydrogen is
H II.  In the current view,  
the intergalactic medium is not uniform -- it has clumped into 
sheets and filaments that we call the Ly$\alpha$ forest. 
Several authors have used high resolution data to put limits on 
any diffuse intercloud medium (Steidel \& Sargent 1987; 
Webb, Barcons, Carswell \& Parnell 1992; Giallongo et al. 1994, 
Songaila et al. 1999).  

\begin{figure}[htb]
\vspace{5pc}
\centerline{\psfig{figure=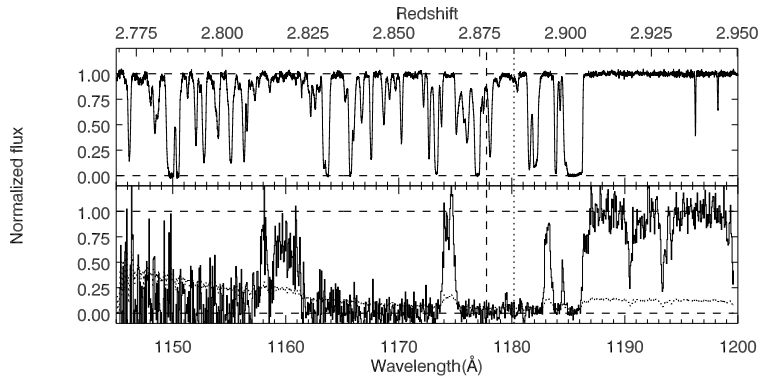,width=10cm,angle=0}}
\caption{Spectra of HE~2347--4342 showing variation of HI/He II at
$z\sim2.8$.
Top: Keck-HIRES  spectrum of H I Ly$\alpha$.
Bottom: {\it HST}/STIS FUV MAMA G140M spectrum of He II.
In both  panels, the vertical  dashed
line at  $\sim  1178~\mbox{\AA}$   corresponds to $z     = 2.877$,
redwards of which  the spectra  are likely  to be  affected by the
known $z_{abs}  \simeq  z_{em}$ systems.  From Smette et al. (2000).
}
\end{figure}

Direct information on the ionization of the intergalactic gas may be
derived from observing the neutral and singly ionized helium
Ly$\alpha$ lines at $\lambda$584 and $\lambda$304, respectively.
Tripp, Green \& Bechtold (1990) used IUE to put an upper limit on 
He I in the IGM using the lack of a depression at $\lambda$584 in a
high redshift quasar.  However, if the quasar radiation field is
ionizing the IGM, most helium is expected to be He II or He III. 

He II $\lambda$304 is difficult to observe because it is so far in
the far ultraviolet.  The redshift of the quasar must be at least 3 to
redshift this feature into the HST range, and at least 2 to redshift
the feature into the HUT/FUSE range.  Most $z=2-3$ quasars have optically
thick Lyman limits at other redshifts which wipe out the continuum 
so that $\lambda$304 is not observable.  To observe He II, one must survey
many quasars to find the one or two which happen to be clear of
Lyman limit absorption.  

Jakobsen et al. (1994) succeeded in observing the He II $\lambda$304 
region in the $z=3.2$ quasar Q 0302-003 with FOC on HST and found a sharp
cut-off in the continuum flux -- implying $\tau_{He II} > 3$.  Tytler et al.
(1995) used FOS to observe a second $z=3$ quasar, PKS 1935-692 and
also saw that He II was optically thick.  Remarkably, Davidson et al. 1996
used HUT to observe HS1700+6416, and found that He II was not completely
absorbed at $z=2.4$, $\tau_{He II} = 1$.  This would imply that 
the He II IGM opacity changed abruptly at $z\sim2.7$.  Subsequently, 
Reimers et al. (1997) used GHRS on HST to observe a fourth quasar, 
HE 2347-4342, and found that at $z\sim2.8$ the He II ionization is 
patchy -- that is, there is significant variation in H I / He II.
Recent reobservation of these object with STIS (Heap et al. 2000; 
Smette et al. 2001) and FUSE (Kriss et al. 2001) 
confirm this -- see Figure 12. 

More information about the ionization state of the Ly$\alpha$ forest
at high redshift 
was found when weak C IV, Si IV and O VI absorption lines 
were detected for Ly$\alpha$
forest lines (Songaila \& Cowie 1996; 
Sargent 1996; Figure 13).  
Remarkably, about 75\% of the Ly$\alpha$ lines
with $N(HI)>10^{14.5}$ cm$^{-2}$ show C IV absorption.  The inferred
metallicity is low, a factor  100 -- 1000 less than solar.  Whether
there is a change in metallicity with redshift is a difficult question
to answer, owing to the large uncertainty in the ionization corrections.
Also it is not clear whether the Ly$\alpha$ lines with lower columns
have similar metallicities or not (Ellison et al. 2000; Schaye et al. 2000).
Nonetheless, the discovery of metals associated with such low column density
absorption means that some sort of chemical enrichment of the IGM 
has taken place by $z=3$.

\begin{figure}[htb]
\centerline{\psfig{figure=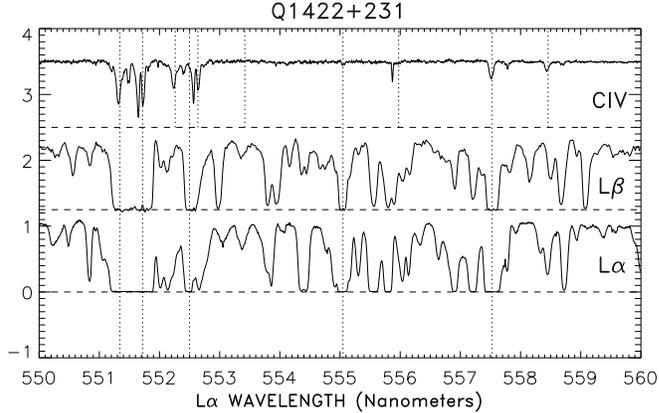,width=10cm,angle=-90}}
\caption{A 100\AA portion of the spectrum of Q1422+231 (bottom) with the
corresponding Ly$\beta$ (middle) and C~IV $\lambda 1548 \AA$ (top)
shifted in wavelength so they lie above Ly$\alpha$.  Dotted vertical
lines indicate clouds that are saturated in both  Ly$\beta$ and 
Ly$\alpha$.  Short dotted lines show the position of C~IV $\lambda 
1550\AA$ corresponding to these clouds. 
From Songaila and Cowie (1996).}
\end{figure}

At low redshift, O VI absorption may be common as well (Bergeron et al. 1994;
Burles \& Tytler 1996; Verner, Tytler \& Barthel  1994;
Dav{\' e} et al. 1998; Tripp, Savage \& Jenkins 2000; Tripp \& Savage 2000;
Reimers et al. 2001; Figure 14) and the warm gas which produces O VI may be an 
important component of the baryons. 
The low redshift systems may not be simply photoionized by the ambient UV
background, however, but may represent shock heated gas associated
with infalling material in groups and clusters of galaxies.

\begin{figure}
\centerline{\psfig{figure=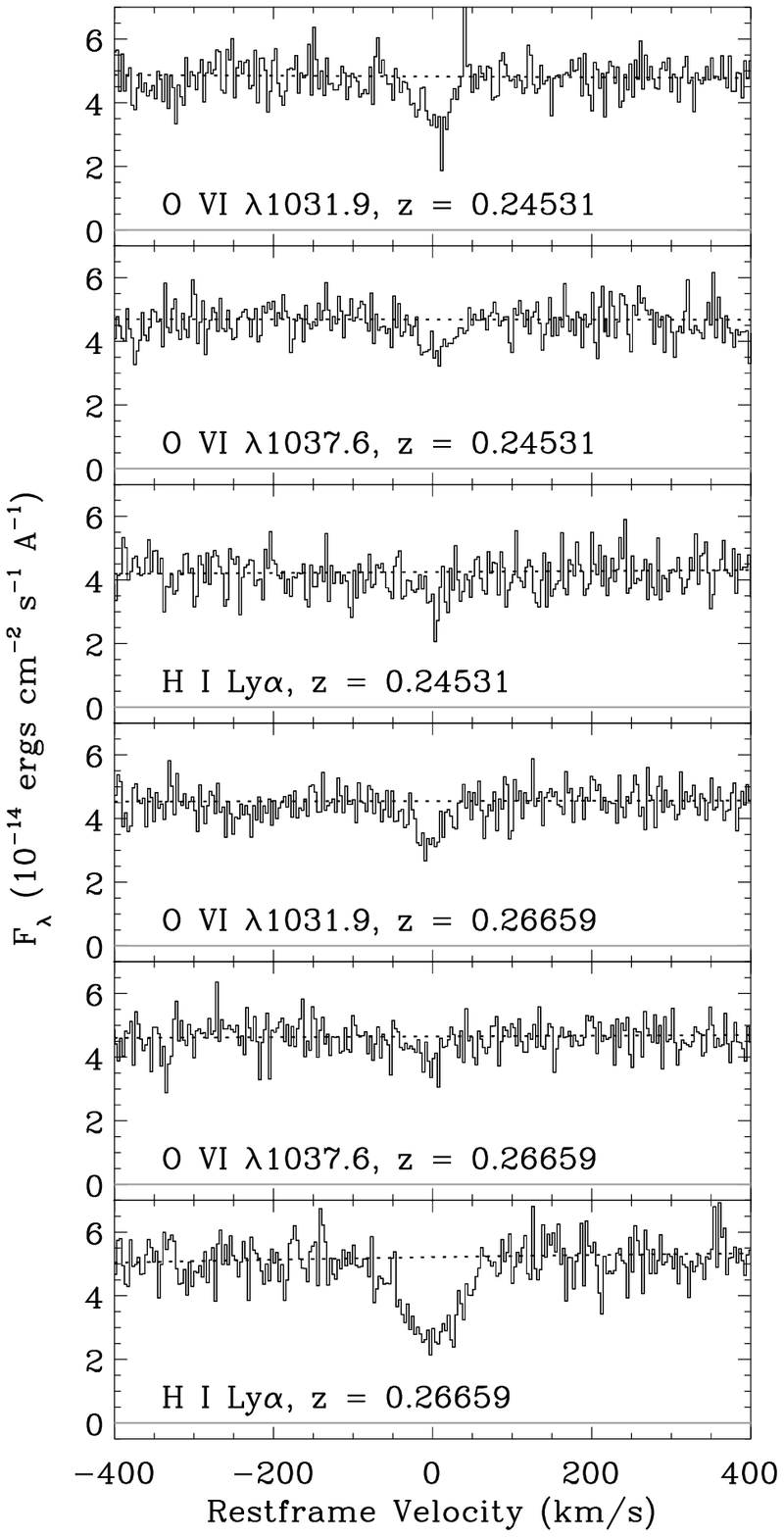,width=12cm,angle=0}}
\caption{Portion of the STIS E140M spectrum of H1821+643 showing
O VI absorption at $z(abs)$=0.24531 (upper three panels) and 
at $z(abs)$ = 0.26659 (lower three panels), plotted versus restframe 
velocity.
From Tripp, Savage \& Jenkins (2000).}
\end{figure}

After this review was completed,  Becker et al. (2001) presented
spectra of the first z$\sim$6 quasars, discovered in the Sloan Survey (see 
http://www.sdss.org).
The very strong Ly$\alpha$ absorption in the highest redshift quasar now known, 
at $z=6.28$, suggests that the IGM was reionized at $z\sim 6$ (Figure 15).  
This relatively low redshift for reionization has interesting 
implications for formation scenarios for galaxies, and strategies for
deep surveys with $NGST$.  Also, since early reionization would have erased
more of the cosmic microwave background 
anisotropies (Ostriker \& Vishniac 1986; Hu \& White 1997; 
Liddle \& Lyth 2000 and references therein), 
reionization at $z\sim$6 means that more cosmological information may potentially be derived from satellite experiments such as MAP and Planck.
 
\subsection{Simulations of the IGM}

\begin{figure}
\centerline{\psfig{figure=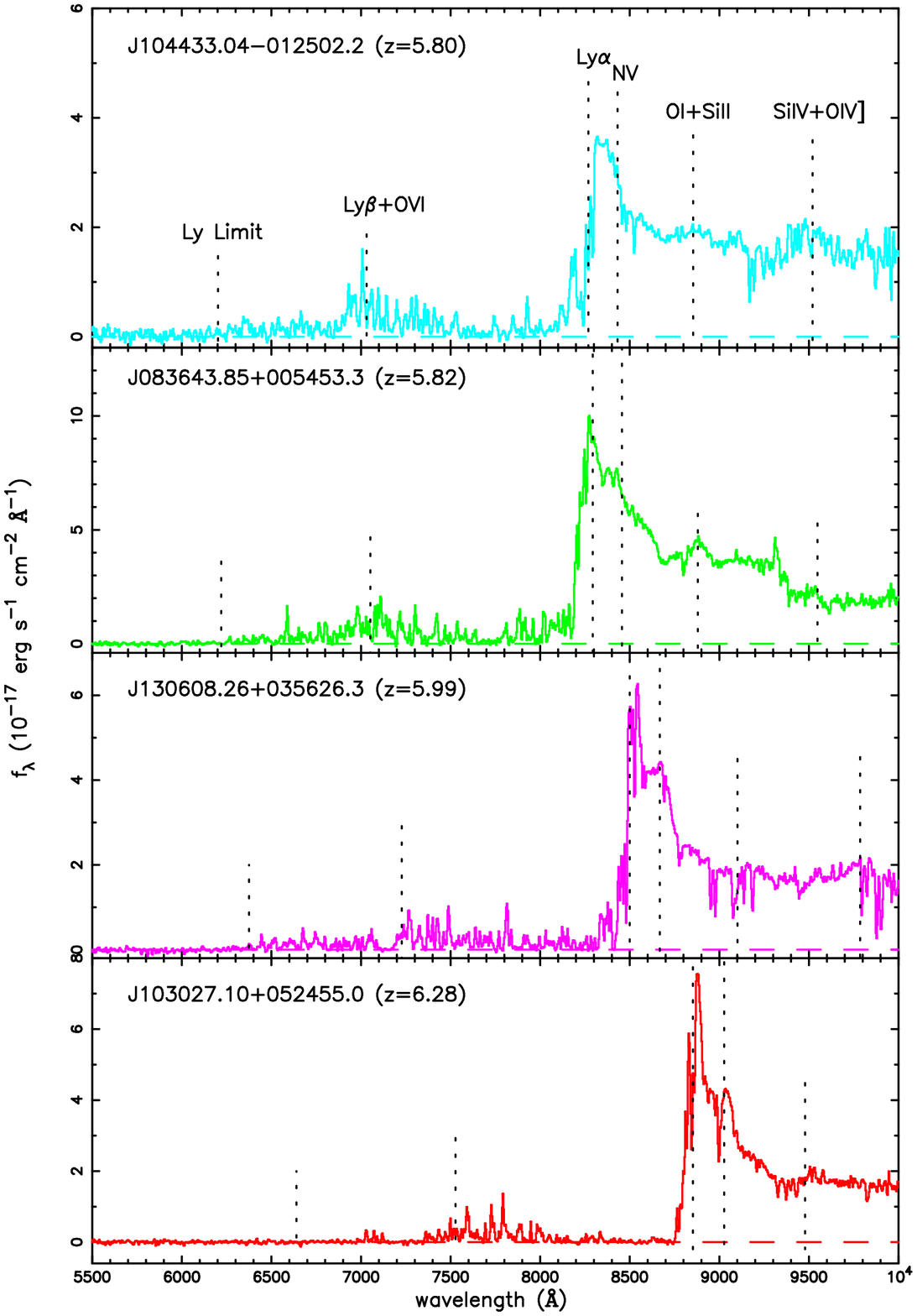,width=10cm,angle=0}}
\caption{Optical spectra of $z>5,8$ quasars observed with Keck/ESI, in the observed frame.  The strong absorption in the $z=6.28$ quasar suggests that
reionization of the intergalactic medium took place at $z\sim$6.
From Becker et al. (2001).}
\end{figure}

The first hydrodynamic cosmological simulations with
sufficient resolution and dynamic range to model
the intergalactic gas showed that the Ly$\alpha$ forest 
is a natural consequence of the growth of structure in
the CDM hierarchical gravitational collapse scenario.  All models 
were remarkably successful in reproducing the
basic observations
(Cen et al. 1994; Zhang, Anninos \& Norman 1995; 
Hernquist et al. 1996; Miralda-Escude, Cen, Ostriker \& Rauch 1996;
Weinberg, Hernquist \& Katz 1997; Wadsley \& Bond 1996; 
Theuns et al. 1998; Haehnelt \& Steinmetz 1998), including the 
the column density distribution and b-value distribution.
There were few free parameters -- an essential one
being the assumed UV radiation field which photoionizes the clouds. 

At high redshifts, $z\sim2-5$, the 
forest is produced by gas with low overdensities, $\rho/\bar{\rho} \le$ 10, 
photoionized by an diffuse ultraviolet background radiation
field. Virtually
all the baryons collapse into structures which can be identified
with observable lines -- there is very little diffusely distributed
gas, or Gunn-Peterson effect. 
The high column density lines ($N_{H I} > 10^{17}$ cm$^{-2}$) are identified 
with high density regions at the intersections of sheets and filaments,
where the density is high and star-formation is presumed to have started.
The low column density lines ($N_{H I} < 10^{12}-10^{15}$ cm$^{-2}$) 
arise from a range of physical conditions and scales, at different
stages of collapse -- including warm gas in filaments and sheets, 
cool gas near shocks, and underdense regions megaparsecs across (voids).  

Since the overdensities producing the Ly$\alpha$ forest lines are 
only close to linear, they can also be approximated with
semi-analytical models (McGill 1990;
Bi \& Davidson 1997; Hui, Gnedin \& Zhang 1997).  These are easier to
calculate than the full 3-dimensional models.
  
The success of the models by different groups using different detailed 
cosmological assumptions meant that 
the observed properties of the Ly$\alpha$ forest clouds that  
were being used for comparison depended only on the assumptions that
are common to all models, that is, the forest provides little 
leverage to distinguish among them (see Weinberg et al. 1998 and 
references therein, also Machacek et al. 2000).  
However, the forest observations do 
allow one to measure the mass power spectrum, $P(k)$, of fluctuations
(Croft et al. 1999), the mass density parameter $\Omega _M$ (Weinberg et al. 
1999) and the spectrum of the UV radiation field which turned out to 
be surprisingly soft (Zhang et al. 1997; Rauch et al. 1997; Zhang et al. 1998).
  
At $z<2$ the models of the intergalactic medium must take into 
account that an increasing fraction of the baryon mass is
participating in star-formation (that is, 
has collapsed into galaxies), or 
has fallen into the cores of clusters of galaxies and been heated to
X-ray temperatures. Details of
gas dynamics, radiative cooling, and $``$feedback" or galactic winds
become important.  One salient feature at low redshift, 
the change in dN/dz evolution at $z\le$2 is easily
reproduced by a modest change in J$\nu$, expected from the changing
luminosity function of quasars and starburst galaxies.  Implications derived
from simulations and a discussion of the observations is given by
Dav{\' e} et al. (1999).
 
\section{Damped Lyman Alpha Absorbers and other Metal-line Absorbers}

The look-back time for quasar absorbers is comparable to the age of
the oldest stars in the Milky Way, so we expect that quasar absorbers
will arise in gas associated with 
galaxies in their earliest stages of formation.
The damped Ly$\alpha$ absorbers have proven particularly valuable
for look-back studies, since the strong absorption of Ly$\alpha$
may be detected with moderate spectral resolution data,
and the damping profile allows an accurate measurement of $N(HI)$,
a first step in metallicity studies.  Art Wolfe and collaborators
have undertaken extensive surveys to find damped Ly$\alpha$ absorbers
and study their properties 
(Wolfe, Turnshek, Smith \& Cohen 1986;
Turnshek et al. 1989;
Lanzetta et al. 1991;
Wolfe, Turnshek, Lanzetta \& Lu 1993;
Lu, Wolfe, Turnshek \& Lanzetta 1993;
Lu \& Wolfe 1994;
Lanzetta, Wolfe, \& Turnshek 1995;
Storrie-Lombardi \& Wolfe 2000). 
Other metal absorption surveys include
those for Mg II (Lanzetta, Wolfe \& Turnshek 1987;
Sargent, Steidel \& Boksenberg 1988;
Barthel, Tytler \& Thomson 1990;
Steidel \& Sargent 1992;
Aldcroft, Bechtold \& Elvis 1994;
Churchill, Rigby, Charleton \& Vogt 1999), 
C IV (Sargent, Steidel \& Boksenberg 1989; Foltz et al. 1986) 
and Lyman limits (Tytler 1982;
Bechtold et al. 1984;
Sargent et al. 1989; Stengler-Larrea et al. 1995).   
A heterogeneous catalog of absorbers from the literature, 
which is nonetheless useful for
some purposes, has been assembled by York et al. (1991).

\subsection{The Mean Free Path for Absorption and its Evolution}

The evolution of the number of 
metal absorbers per redshift can be parameterized by $\gamma$, 
for the Ly$\alpha$ forest clouds.  In contrast to the steep evolution seen for 
the low column forest clouds, the Lyman limit absorbers and damped Ly
$\alpha$ absorbers are consistent with little if any evolution
from $z=0$ to $z=5$.   
Figure 16 shows the evolution of the number of absorbers per redshift
versus redshift.  

Assuming
that the density of absorbers is given by the galaxy luminosity function,
and that the cross-section for absorption, $\sigma_o$, 
scales with luminosity, it is straight-forward to 
calculate how big the gaseous extent of a typical (say L$^{*}$) galaxy 
must be in order to account for the number of absorbers 
seen in quasar spectra (e.g. Burbidge et al. 1977).  
The relatively large sizes led to the notion that the quasar absorbers
arise in huge gaseous halos. While at low redshift this may be 
a valid interpretation, at high redshift, the fact that the 
number density of absorbers is changing must be included.
In light of other results which suggest
that large galaxies assembled out of numerous protogalactic fragments
at $z<3$ (reviewed by Dickinsin at this school), 
the statistics probably indicate 
instead that mergers of PGFs have taken place.  

\begin{figure}
\centerline{\psfig{figure=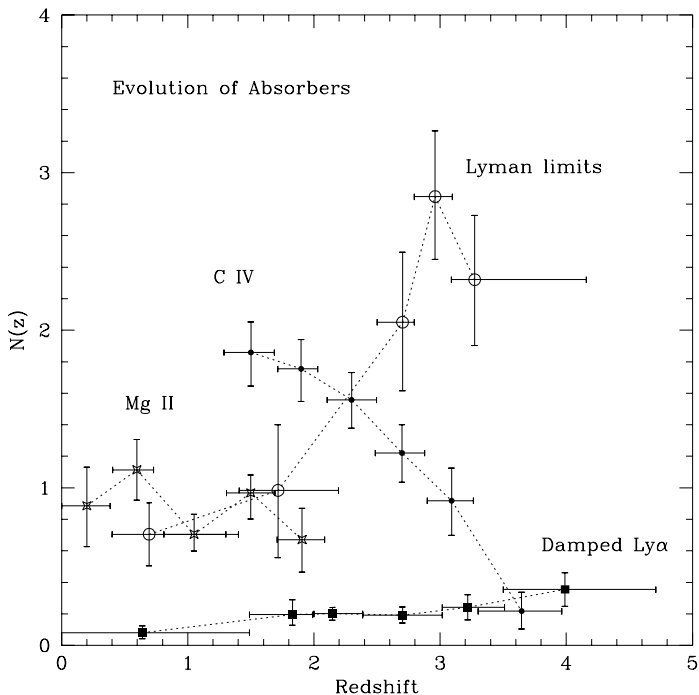,width=10cm,angle=0}}
\caption{Number of absorbers per unit redshift, N(z), as 
a function of redshift, for damped Ly$\alpha$ systems (filled square), 
Mg II absorbers (cross), C IV absorbers (filled circles)
 and Lyman limits (open circles). Data from Storrie-Lombardi \& Wolfe (2001),
Stengler-Larrea et al. (1995), and  the compilation of York et al. (1991).}
\end{figure}

The evolution of the damped Ly$\alpha$ absorbers is of interest, because
at low redshift they appear to arise in normal galaxies, and
at high redshift, they appear to contain a large fraction of the
available baryon density of the Universe (Lanzetta et al. 1991;
Rao, Turnshek \& Briggs 1995; Rao \& Turnshek 1999;
Storrie-Lombardi \& Wolfe 2000 and references therein).

\begin{figure}
\centerline{\psfig{figure=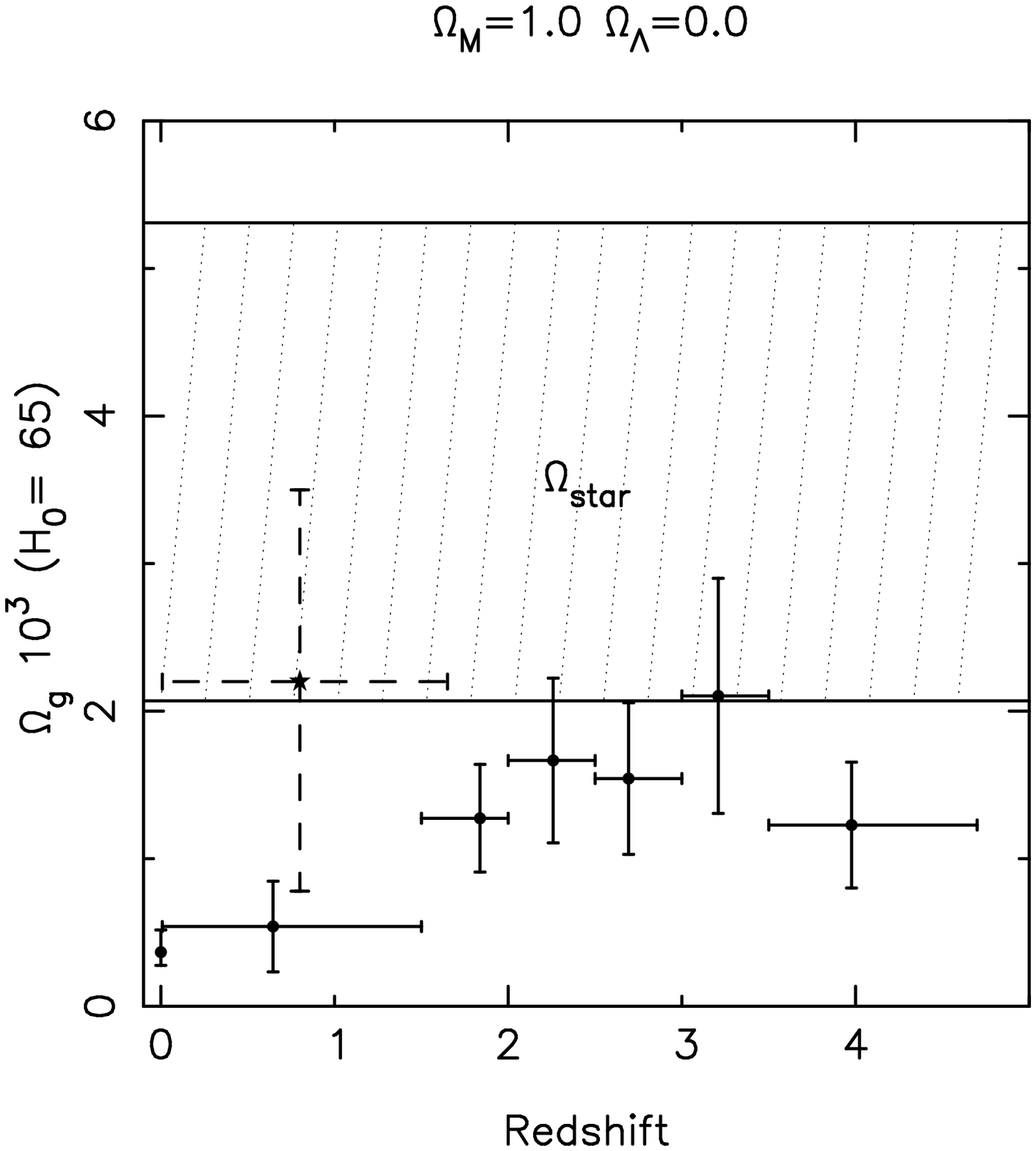,width=10cm,angle=0}}
\caption{
For $\Omega _{matter}=1$, $\Omega _{\Lambda}$ = 0, 
the comoving mass density in neutral gas contributed
by damped Ly$\alpha$ absorbers, $\Omega _{g}(z)$, compared to 
$\Omega _{star}$, the comoving mass density of stars.
The dashed lines are a determination $\Omega _{g}(z)$ from the survey of  
Rao \& Turnshek (2000) for $z < 1.65$.
From Storrie-Lombardi \& Wolfe (2001).}
\end{figure}

The column density distribution of the damped Ly$\alpha$ 
may be integrated to estimate the comoving mass density of the gas in
units of the current critical density,

\begin{equation}
\Omega_g(z) = {\frac{H_{0}}{{c}}}{\frac{{\mu}m_{H}}{\rho_{crit}}}{\frac{\sum_{i}N_{i}({HI})}{\Delta X(z)}}.
\end{equation}

\noindent
Here $\mu$ is
the mean particle mass
per $m_{H}$ where the latter is the mass of
the H atom, $\rho_{crit}$ is the current critical mass density,
and the sum is over damped Ly$\alpha$ systems in the sample surveyed over
distance X(z), given by Equation 4.15.
$\Omega_g(z)$ is somewhat sensitive to the column density of the highest
column density absorber in the sample, and there is a strong selection
{\it against} the high column density absorbers in UV studies because of
reddening.  However, the results, shown in Figure 17, are interesting.  
The evolution of $\Omega_g$ as measured by quasar absorbers is compared
to the mass density of stars in nearby galaxies, as estimated by
Fukugita, Hogan \& Peebles (1998).  The point at $z=0$ is from a 21cm 
emission survey of local galaxies (Zwaan et al. 1997).  The suggestion
is that the $z\sim3$ damped Ly$\alpha$ gas has been converted into
the stars we see in local galaxies today. 

\subsection{The Damped Lyman Alpha Absorbers: Metallicity and Dust}

The aim of abundance studies of quasar absorbers were 
described by Trimble (1995) as an investigation of Z(Z(Z)) -- 
that is, the abundance as a function of scale height as a function 
of redshift. 
The damped Ly$\alpha$ absorbers are the ideal absorbers for
abundance studies since they have high total columns, so 
that weak transitions are detectable of rare elements 
on the linear part of the curve of growth.

A serious complication of all abundance studies based on 
UV absorption is that the elements of interest suffer 
depletion -- they freeze out onto dust grains, so 
the gas phase column density measured by the absorption lines
is less than the actual abundance of the gas (see review by
Jenkins 1995).  In the Milky Way, the 
refractory elements (Si, Fe, Ca, Ti, Mn, Al 
and Ni) show depletion factors  of 10-1000, and the exact 
pattern of depletion depends on the shock history of the gas.
For C, N and O, the depletions are factors of 2 to 3.     
S, Zn and Ar are relatively undepleted, and therefore can be used to 
measure metallicities.

Pettini and collaborators did a comprehensive survey of Zn II in
damped Ly$\alpha$ absorbers, using the William Herschel Telescope 
on La Palma (Pettini, Smith, Hunstead \& King 1994, 1997;
Pettini, King, Smith \& Hunstead 1997;
Pettini \& Bowen 1997;
Pettini, Ellison, Steidel \& Bowen 1999;
Pettini et al. 2000;  see also Meyer, Welty \& York 1989; 
Meyer \& Roth 1990;
Meyer \& York 1992;
Sembach, Steidel, Macke \& Meyer 1995;
Meyer, Lanzetta \& Wolfe 1995).
The result was that the zinc to hydrogen abundance is
about 1/10-1/50 solar, with a large scatter.  They are generally
more metal poor than Milky Way disk stars of the same age as the
lookback time to the absorbers (Figure 18).   The quasar absorbers are less
metal-rich than thin or thick disk stars, but not as metal poor
as halo stars or globular clusters in our Galaxy (Pettini et al. 1997; 
Figure 19).  
Zinc is an iron group element, so even though it is difficult
to measure [Fe/H] directly, the evolution of the [Zn/H] ratio is
probably indicative of evolution of [Fe/H].

\begin{figure}
\centerline{\psfig{figure=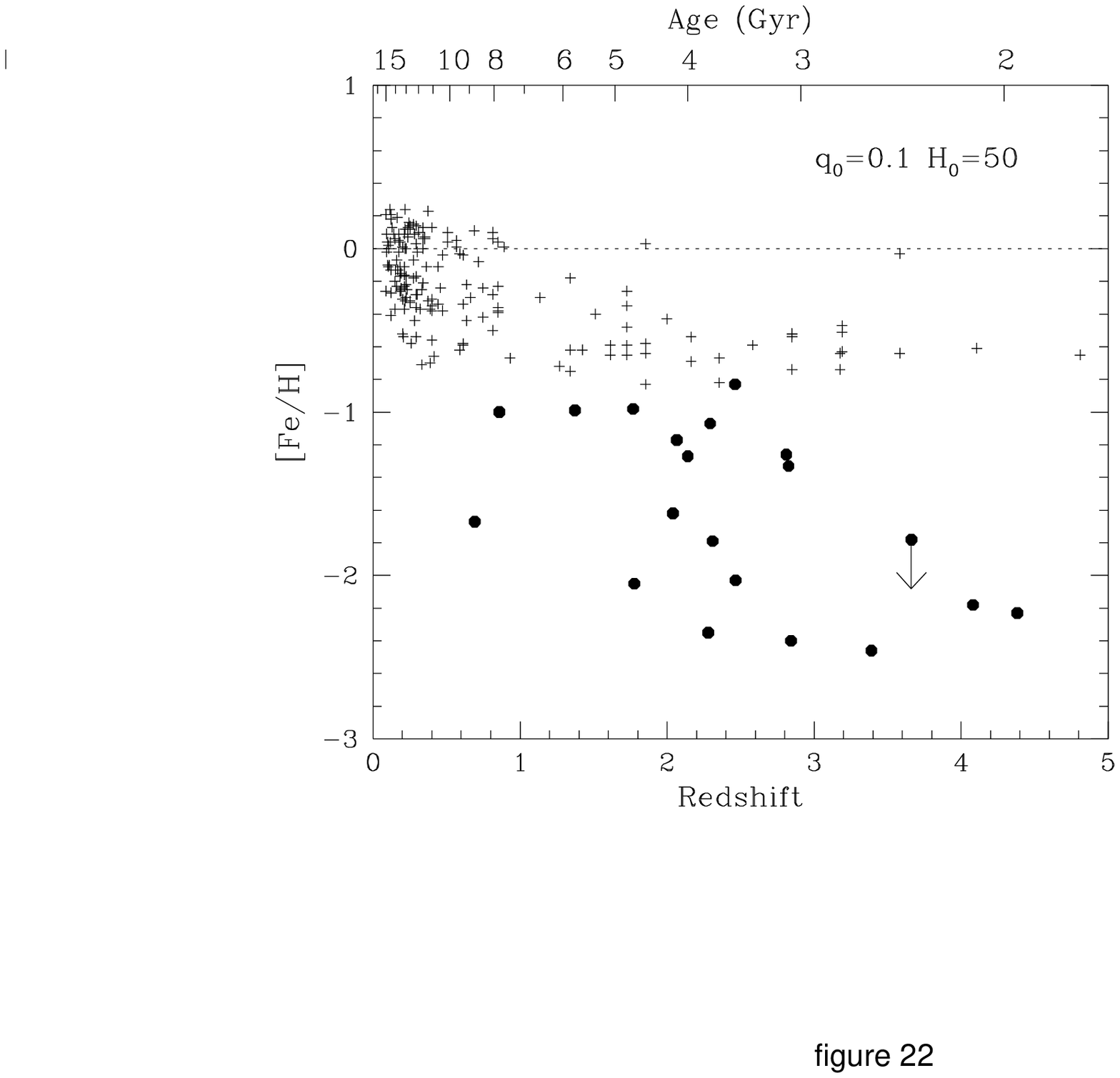,width=14cm,angle=0}}
\caption{ Distribution of [Fe/H] as a function of
redshift for damped Ly$\alpha$ absorbers (solid circles).  
The age-metallicity relation for Galactic disk stars 
(Edvardsson  et al.  (1993) is shown with ``+'' symbols.
The damped Ly$\alpha$ galaxies have
Fe-metallicities in the
range of 1/10 to 1/300 solar, or [Fe/H] between $-1.0$ and $-2.5$.
  From Lu et al. (1996).}
\end{figure}

\begin{figure}
\centerline{\psfig{figure=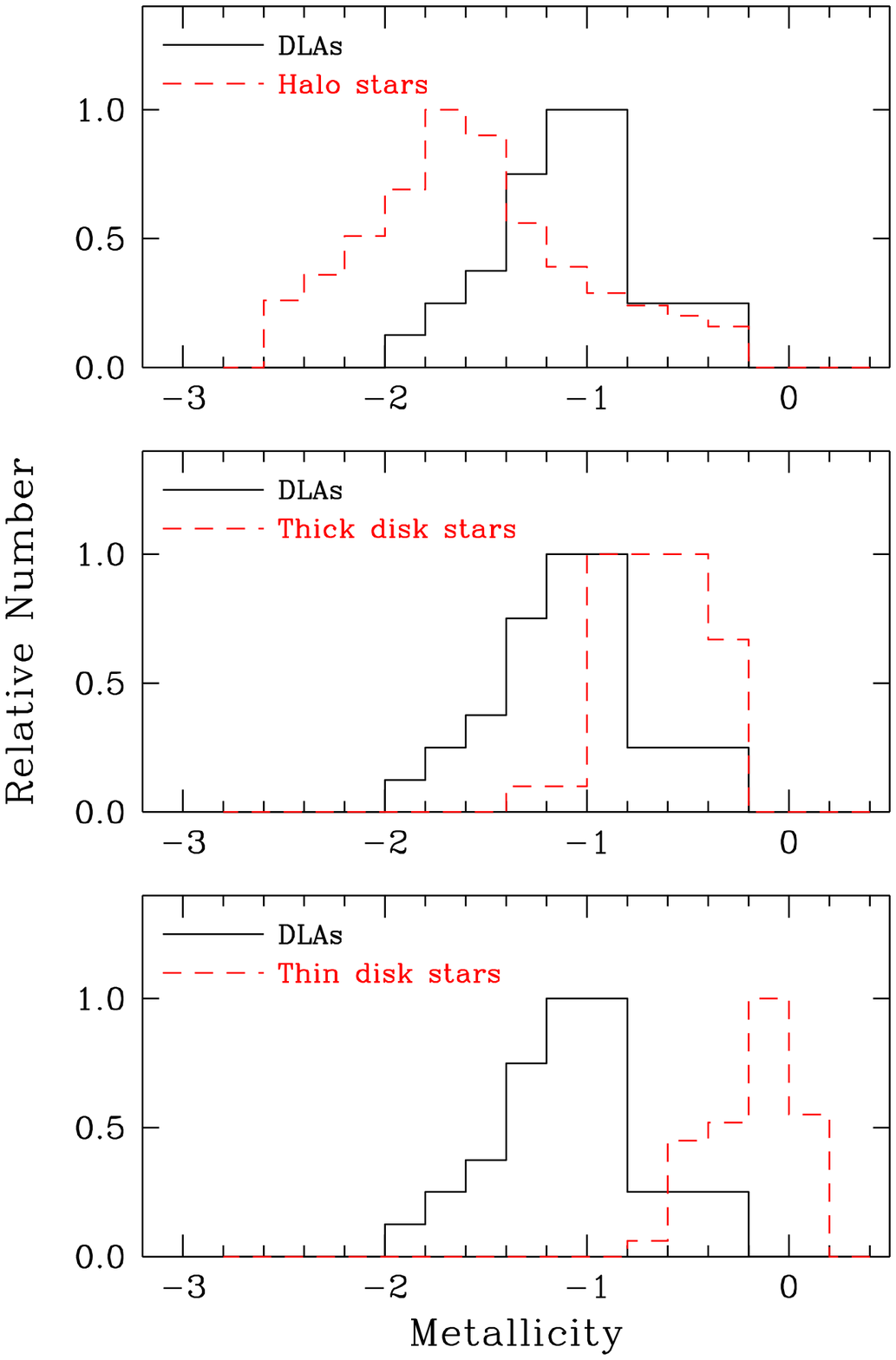,width=12cm,angle=0}}
\caption{Metallicity distributions, normalised to unity,
of damped Ly$\alpha$ systems and of stars belonging to the
disk and halo populations in the Milky Way.
 From Pettini et al. (1997).}
\end{figure}

Keck HIRES data allowed more detailed studies of the abundance
patterns of the elements to be carried out (Prochaska \& Wolfe 1996, 
1997a, 1999; Lu et al. 1996).  The overall result
is that the extent of depletion onto dust grains is small, and
the pattern of iron group to $\alpha$ rich nuclei suggests an
origin similar to Population II Milky Way objects, which are
enriched early by Type II supernovae.  Ionization corrections
may mimic halo $\alpha$ element enrichment, however,
and should be accounted for (Howk \& Sembach 1999). 

The ratio of nitrogen to oxygen is also of interest, since
if whether or not N/O depends on O/H can distinguish between
the $``$primary" or $``$secondary" production of nitrogen.  Primary
production of nitrogen results from the asymptotic branch phase
of intermediate mass stars, which depends on CNO burning of seed 
carbon nuclei dredged up from the interior of the star -- thus
N/O should be independent of the initial metallicity of the star.
If nitrogen is produced by the CNO cycle of main sequence stars 
($``$secondary" production) then it depends on the initial metallicity.
At low metallicity, one expects a large scatter in N/O because of 
the delay between the release of primary nitrogen from intermediate mass 
stars and oxygen from massive stars.
The problem with quasar absorbers is that it is oxygen and nitrogen are both 
difficult to measure since the lines are saturated or in the Ly$\alpha$
forest.  The preliminary results suggest that there is a large
scatter at low metallicity, larger than that seen in low metallicity
dwarf galaxies locally (Pettini, Lipman \& Hunstead 1995; Lu, Sargent \& Barlow 1998).

A summary of the heavy element abundance studies of quasar 
absorbers is shown in Figure 20.  

\begin{figure}
\centerline{\psfig{figure=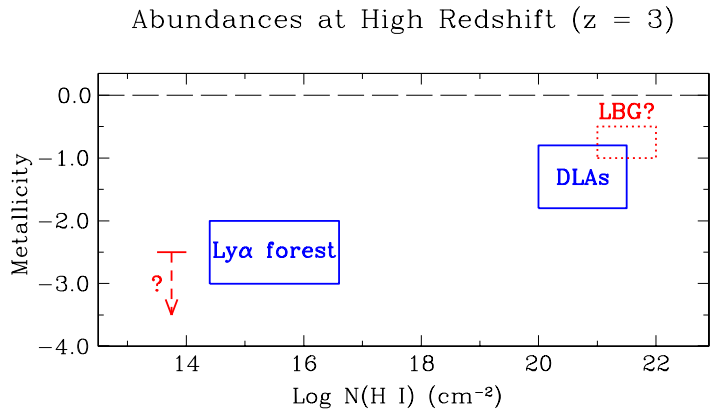,width=14cm,angle=-90}}
\caption{Summary of abundances relative to solar at high
redshift.  $N(HI)$ is the column density of neutral hydrogen,
shown for the Ly$\alpha$ forest, damped Ly$\alpha$ absorbers 
(DLAs) and Lyman break galaxies. From Pettini (1999).}
\end{figure}

\subsection{The Damped Lyman Alpha Absorbers: Kinematics}

\begin{figure}
\centerline{\psfig{figure=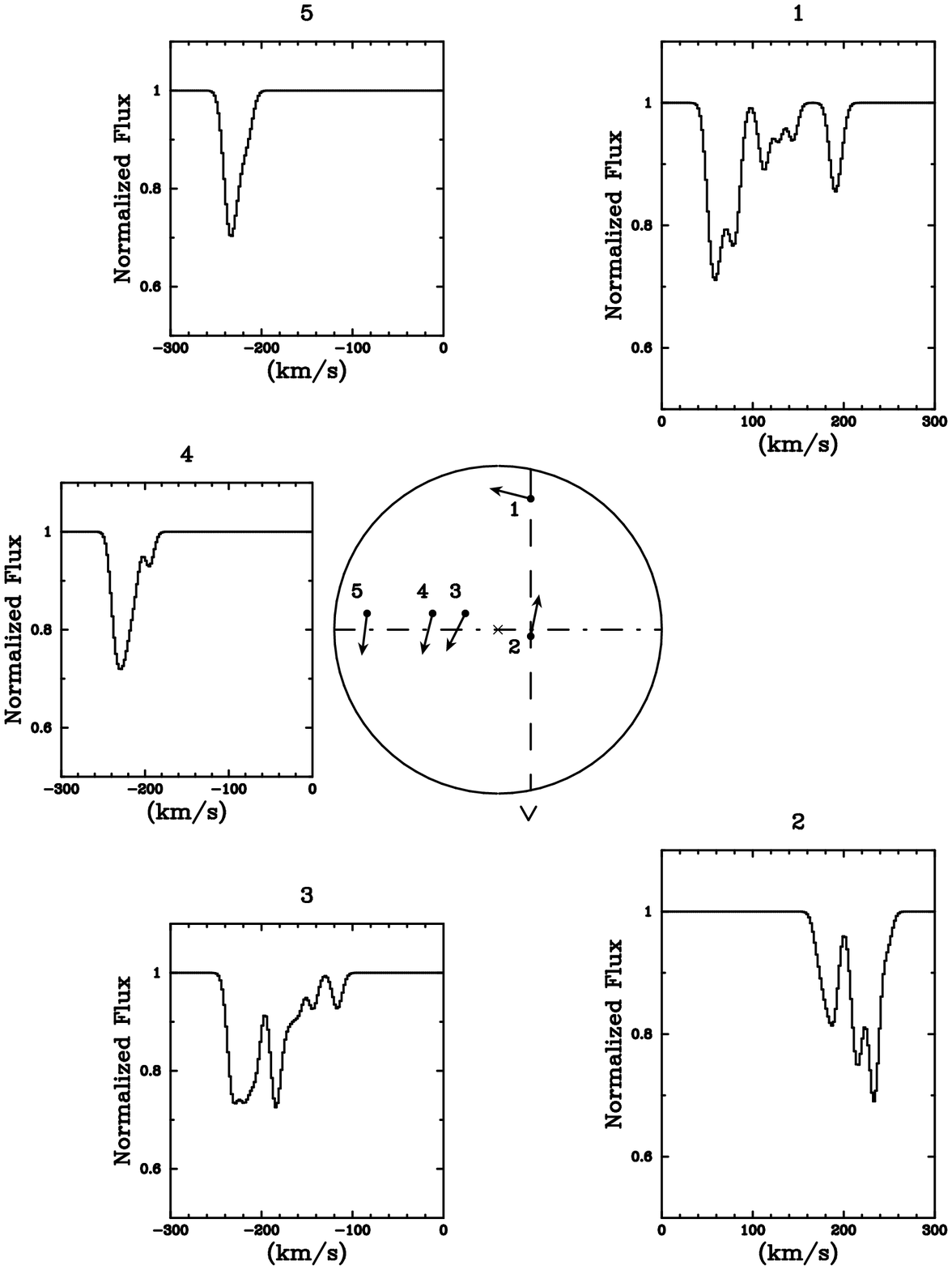,width=10cm,angle=0}}
\caption{Expected absorption line profiles from a rotating disk of 
gas.  The center circle represents an inclined disk (top view)
rotating counter clockwise with rotation speed $v_{rot} = 250$ km s$^{-1}$.
The solid dots represent the intersection points for 5 different
sightlines with the midplane of the disk.
The solid arrows indicate the direction of the rotation vector.
The sightlines are inclined by
70 $\deg$ with respect to the normal of the disk and yield the profiles
labeled 1$-$5. From Prochaska and Wolfe (1997).}
\end{figure}

The absorption profiles of metal lines which aren't too badly saturated
give information on the kinematics of the gas in high redshift
galaxies.  If the galaxy has a rotating disk with a radial gradient of
decreasing gas density with increasing radius, then the 
absorption profile of a typical line-of-sight has
a distinctive signature (Weisheit 1978), with the strongest absorption 
velocity component on the extreme red or blue side of the profile. 

Prochaska \& Wolfe (1997b, 1998) showed that indeed many 
damped Ly$\alpha$ systems have 
the expected absorption profile if the absorbing velocity clouds are 
in a rotating thick disk. A detailed analysis of one system 
at $z=3.15$ with a detected emission line galaxy also was consistent
with the expectations of a rotating disk (Djorgovski, Pahre, 
Bechtold \& Elston 1996; Lu, Sargent \& Barlow 1997).  Prochaska $\&$ 
Wolfe further compared the profiles observed in a large sample of 
about 30 systems, with what you would expect to see if the 
quasar sight lines randomly sampled clouds with various geometries
and kinematics, including isothermal halos, spherical distributions 
undergoing collapse, and thick and thin rotating disks.  
They argued that the distribution of line profiles favored a model 
where 
the damped Ly$\alpha$ absorbers originate in rapidly rotating disks, 
with maximum velocities, $\Delta$v $\sim$ 100-300 km s$^{-1}$ (Figure 21).  
Moreover, the average size and $\Delta$v of the disks
was inferred to be larger than what was comfortably predicted by
semi-analytical models for the evolution of galaxy disks in
the cold-dark-matter hierarchical collapse picture, as described
for example by Kauffmann (1996).  However, Haehnelt, Steinmetz \& Rauch 
(1998; Figure 22)
showed that collapsing protogalactic fragments in their n-body/hydro 
simulations
also rotate, and could reproduce the distribution of line
profiles seen by Prochaska $\&$ Wolfe, without the need for large,
rapidly rotating disks.  The full-up simulations include non-linear
effects which aren't included in the linear Press-Schechter 
theory.  For merging protogalactic clumps, unlike the case for
Ly$\alpha$ forest clouds, it is crucial to include non-linear effects
in the models. 

\begin{figure}
\centerline{\psfig{figure=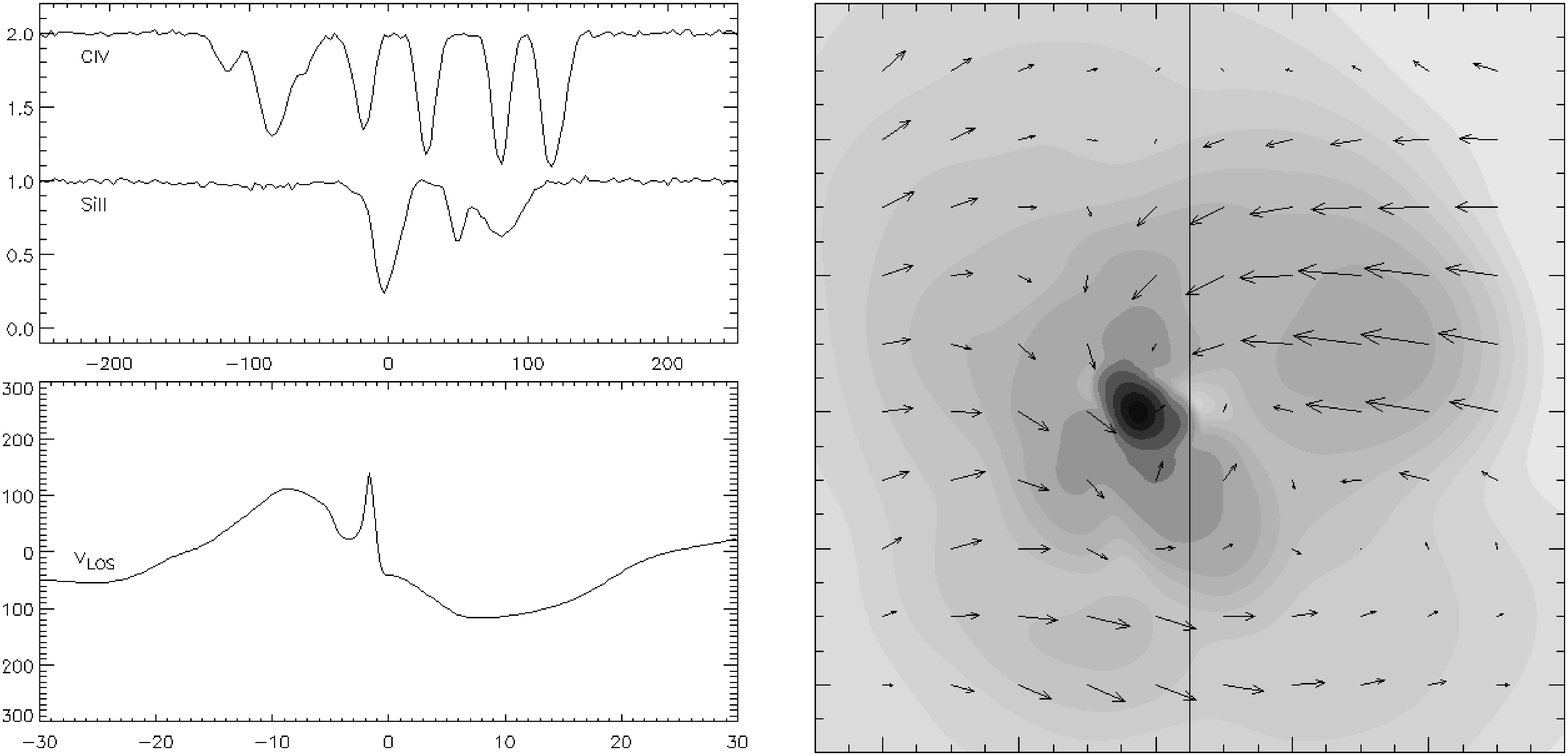,width=10cm,angle=0}}
\caption{ Velocity field of a collapsing protogalactic fragment 
which shows the same absorption line profile as a larger,
rotating disk.  From Steinmetz (2000).}
\end{figure}
 
At lower redshift, Churchill, Steidel \& Vogt (1996) and 
Churchill et al. (1999) have studied the  
kinematics of Mg II absorbers.  The distribution of  
line profiles favors an origin in galaxies with large rotating disks
and an infalling gaseous halo.  These results are consistent
with the direct imaging of the absorbing 
galaxies, discussed below, which suggest that the $z<1$ absorbers
are relatively mature galaxies, similar to ones seen locally.
 
\subsection{Absorption by Molecules}

Since high redshift quasar absorbers are pointers to gas-rich galaxies 
with active star-formation, they undoubtedly also contain molecular gas,
mostly H$_2$.
Warm H$_2$ can be detected through IR emission, but most H$_2$ is probably
contained in a cold phase, measurable by ultraviolet Lyman and Werner
absorption (Black $\&$ Dalgarno 1976).
Shortly after the detection of H$_2$ absorption in local
interstellar clouds with Copernicus in the 1970's (Spitzer et al. 1973),
searches for H$_2$ in quasar absorbers began (Aaronson, Black $\&$
McKee 1974;  Carlson 1974).
The UV H$_2$ lines have a distinctive pattern which is straightforward
to distinquish from the Ly$\alpha$ forest (Figure 23).

\begin{figure}
\centerline{\psfig{figure=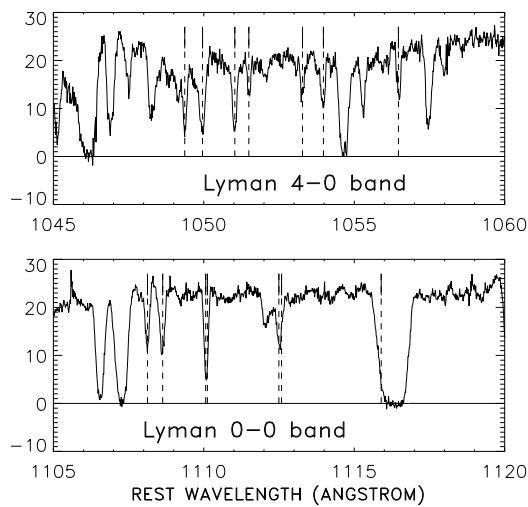,width=10cm,angle=90}}
\caption{Spectra of the 0-0 and 4-0 vibration transitions of the 
Lyman series of molecular hydrogen at $z=2.811$ in PKS 0528-250.  The 
spectrum is shown in the rest frame, and the dasehed lines indicate 
the expected wavelengths of the $J\le 2$ transitions.  
From Cowie $\&$ Songaila (1995).}
\end{figure}

However, evidence for molecular gas or dust in quasar absorbers has
been scant (
Black, Chaffee \& Foltz 1987;
Foltz, Chaffee \& Black 1988;
Lanzetta, Wolfe \& Turnshek 1989;
Levshakov, Chaffee, Foltz \& Black 1992; 
Bechtold 1996, 1999, Ge \& Bechtold 1997, 1999; 
Petitjean, Srianand \& Ledoux 2000;
Levshakov et al. 2000;
Ge, Bechtold \& Kulkarni 2001; Figure 24).
The inclusion of quasars for absorption studies that are bright
and blue selects against lines-of-sight which contain dust, and hence
detectable molecular absorption.

For the absorbers with detected H$_2$, the inferred  
molecular fraction 

\begin{equation}
f_{H_2} = { {2 N(H_2)} \over {2 N(H_2) + N(H I)}}
\end{equation}  

\noindent
is $f_{H_2}=$0.07-0.1 whereas the limits for systems where H$_2$ is not
detected is typically $f_{H_2}< 10 ^{-5}$.

\begin{figure}[htb]
\centerline{\psfig{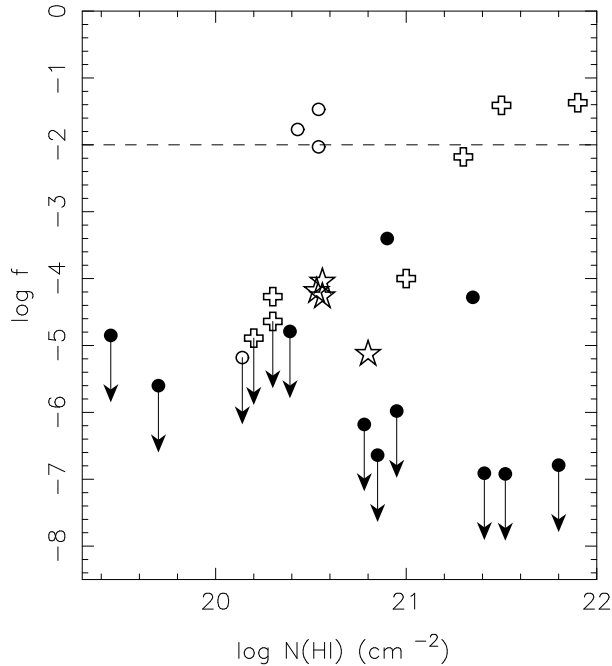}}
\caption{Molecular fraction, $f$~= 2$N$(H$_2$)/(2$N$(H$_2$)+$N$(H~{\sc i})),
versus H~{\sc i} column density. Damped Lyman-$\alpha$ systems
observed at high spectral resolution are indicated by filled circles
(verticle arrows are for upper limits).
Measurements in the Magellanic clouds are indicated by open crosses.
Open circles and open stars are for, respectively, 
lines of sight perpendicular to
the disk of our Galaxy and lines of sight toward the Magellanic clouds.
The horizontal dashed line marks the transition from low to high
molecular fraction in the local Galactic gas ($f$~=~0.01).
From Petitjean, Srianand \& Ledoux (2000).}
\end{figure}

The bimodality of $f$ is
understood as the result of self-shielding.  Once H$_2$ begins
to form, it shields itself from subsequent
photodissociation by ultraviolet radiation, and $f$ increases.

High resolution spectroscopy of the H$_2$ lines give estimates of
excitation temperatures, volume densities (assuming some of the
excitation is collisional), and the ultraviolet radiation field
which can excite or photodissociate the molecules.
The lack of molecules in some absorbers
may be the result of an enhanced ultraviolet
radiation field.  For example, the well-studied absorber of
PHL 957 (Q0100+13) has N(HI) = 2.5$\times$ 10$^{21}$ cm$^{-3}$,
the column density is high enough that H$_2$ should have
been detected (Black, Chaffee $\&$ Foltz 1987).  They suggested that
the lack of H$_2$ in that system resulted from photodissociation
by the ultraviolet radiation field produced by the intense star-formation
in the galaxy within which the absorption features arise.

A related observation is the measurement of the excited fine structure
lines of ions such as Si II, C II and particularly C I, which can
be used to place limits on the temperature of the cosmic microwave
background as a function of redshift (Bahcall \& Wolf 1968;
Meyer et al. 1986;
Songaila et al. 1994;
Lu et al. 1996;
Ge, Bechtold \& Black 1997;
Roth \& Bauer 1999).  For standard expanding
Universes, the microwave background has temperature $T(z) = T_o (1+z)$,
where $T_o=2.7$ K, the value measured today.  
C~I~$^*$ in particular
has a small excitation energy and is well suited for this measurement. 
The problem is that C I can be ionized by photons longward of the Lyman limit
and is rare in redshifted systems;  those showing $H_2$ are generally shielded
from UV radiation adequately to have detectable C I as well.  
Thus far, the few systems where C I has been detected, and the excited
states measured or limited, give $T(z)$ which (unsurprisingly) is 
consistent with the predictions of standard cosmology.
As mentioned above, direct excitation by CMB photons is not the only
source of excitation, and so these estimates of $T(z)$ are really 
upper limits to the cosmic microwave background temperature. 

Molecular absorption in the millimeter (Figure 25) has been detected for 
many different molecules in 
redshift systems at $z=0.2-0.9$ by 
Wiklind and Combes (1994ab, 1995, 1996ab, 1998, 1999), 
Combes \& Wiklind (1997),
and Carilli \& Menten (2000). The millimeter transitions 
can be used probe detailed molecular chemistry. Isotope
ratios such as $^{12}$C/$^{13}$C have been measured; this ratio is
expected to decrease in time
with increasing stellar processing since $^{13}$C is produced in low and
intermediate mass stars whereas $^{12}$C is produced in massive 
stars as well (Wilson \& Matteucci 1992). In one case, 
CO absorption identified the
redshift of the intervening galaxy lensing a background radio
AGN (Wiklind \& Combes 1996a).  
These observations complement the UV studeis, since so far detections
of molecular absorption have been achieved for heavily reddened
objects only, where the optical/UV continuum is too faint for absorption line
spectroscopy.  

\begin{figure}[htb]
\centerline{\psfig{figure=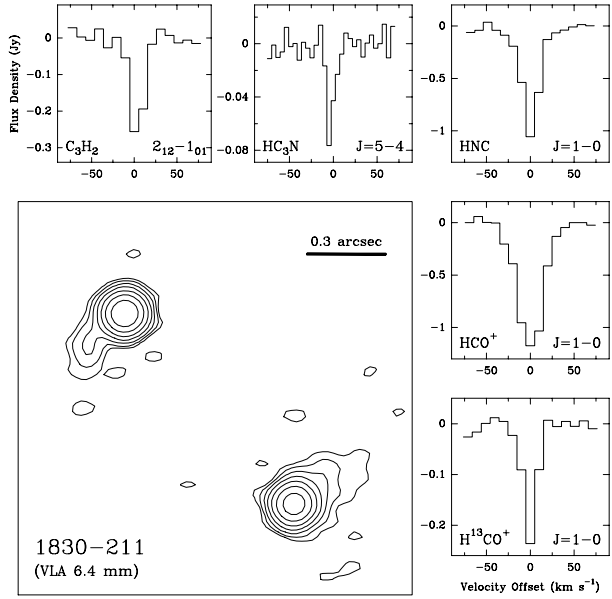,width=11cm,angle=0}}
\caption{Molecular absorption associated with the Einstein ring, 
PKS 1830-211.  Contour plot shows source continuum, spectra 
show molecular absorption by gas in the lensing galaxy toward the southwest
component.  Zero velocity corresponds to 
$z = 0.88582$.  From Carilli \& Menten (2000).}
\end{figure}

Redshifted molecular transitions can also be used to measure the
variation of fundamental physical constants with time, an observational
constraint of interest to certain unified theories
(Varshalovich $\&$ Potekhin 1995, 1996ab; Potekhin et al. (1998); 
Drinkwater et al. 1998).
Specifically, the
wavelengths of H$_2$  are sensitive to
the ratio of electron to proton inertial mass (Thompson 1975) and
observations of H$_2$ at $z=2.8$ by Foltz, Chaffee $\&$ Black (1988) put
an upper limit on the fractional variation of 2.0$\times$ 10$^{-4}$.
Cowie \& Songaila (1995) improved on this by about a factor of 10.

Although all known damped Ly-$\alpha$ absorbers appear to be enriched
with heavy elements (Pettini et al. 1999 and references therein), as
are the quasars used to find them (Ferland et al. 1996, Hamann \& 
Ferland 1999),
someday one may be able to find UV bright objects at high enough
redshift to probe truly primordial H$_2$.  The chemistry of primordial
molecule formation
is important for models of the thermal history of the
early intergalactic medium since H$_2$ cooling controls
the collapse and fragmentation of the first
PGFs.  Observations of primordial H$_2$ would allow interesting measures of
the ultraviolet radiation background, which may
suppress the formation of low-mass objects (Couchman $\&$ Rees 1986;
Efstathiou 1992; Haiman, Rees $\&$ Loeb 1996).

\subsection{Deuterium}

Deuterium was produced throughout the Universe 
about a minute after the Big Bang,  
and has been destroyed by astration ever since (Reeves,
Audouze, Fowler \& Schramm 1973;
Schramm \& Turner 1998).
The primordial ratio of deuterium to hydrogen produced during Big Bang
nucleosynthesis depends on the baryon-to-photon ratio, $\eta$; combined
with the photon density of the cosmic microwave
background radiation (measured precisely with $COBE$), 
a measurement of the primordial D/H therefore
measures the cosmic baryon density, $\Omega_{baryon}$.  Since D/H is destroyed in stars, 
D/H decreases with time; any measurement is a lower limit 
to D/H and hence an upper limit to $\Omega_{baryon}$.  High redshift
measurements of D/H should provide the best limits on $\Omega_{baryon}$,
being closer to the primordial, big-bang value.  A decline of D/H
with cosmic formation time would be a satisfying confirmation of 
the big bang nucleosynthesis theory. 

Deuterium in quasar absorbers has been searched for by measuring the 
deuterium Ly$\alpha$ line, located 
81 km s$^{-1}$ blueward of hydrogen Ly$\alpha$ (
Chaffee, Weymann, Strittmatter \& Latham 1983;
Webb, Carswell, Irwin \& Penston 1991;
Songaila, Cowie, Hogan \& Rugers 1994;
Carswell et al. 1994, 1996;
Wampler et al. 1996;
Tytler, Fan \& Burles 1996;
Rugers \& Hogan 1996ab;
Songaila, Wampler \& Cowie 1997;
Hogan 1998;
Burles \& Tytler 1998abc;
Molaro, Boniracio, Centurion \& Vladilo 1999;
Burles, Kirkman \& Tytler 1999;
O'Meara et al. 2001;
D'Odorico, Dessauges-Zavadsky \& Molaro 2001;
Kirkman et al. 2001;
Tytler, O'Meara, Suzuki \& Lubin 2001).  There are two 
practical difficulties.  First, to have detectable
deuterium given the expected cosmic abundance, the hydrogen column 
density  must be relatively
high, so hydrogen Ly$\alpha$ and the Lyman series  
lines are very saturated, and N(H) is correspondingly
uncertain.  In damped Ly$\alpha$
systems, Ly$\alpha$ is too broad to allow deuterium to be measured at all.
The systems with high enough column to detect deuterium, 
invariably contain metals, so some astration has already 
taken place.  Second, metal absorbers typically have 
velocity substructure on scales of $\Delta v \sim$ 80 km s$^{-1}$ so
one must search a large number of redshift systems to argue statistically 
that one is not seeing merely hydrogen Ly$\alpha$ interlopers (which 
should occur in equal numbers on the blue and red side of Ly$\alpha$).  
On the other hand, for redshifted absorbers, the far ultraviolet lines
of the Lyman series are accessible from the ground, making a larger number
of systems measurable than with HST. 

These difficulties have lead to some debate in the literature about
the D/H ratio in quasar absorbers, and whether or not it is consistent
with the abundances of $^4$He and $^7$Li, also produced by 
big bang nucleosynthesis.  Most of the disagreements were 
the result of curve-of-growth uncertainties in the column of hydrogen, 
and other technical
details, and have since 
been sorted out by the authors.  Figure 26 summarizes the
quasar absorber measurement of D/H  with the predictions from
the Standard Big Bang Nucleosynthesis calculations 
(Burles $\&$ Tytler 1998b), based on ground-based high resolution data 
for absorbers at $z\sim$3.  Absorbers at $z\sim$1 have also been measured
with HST (Webb et al. 1997; Shull et al. 1998).     

\begin{figure}[htb]
\centerline{\psfig{figure=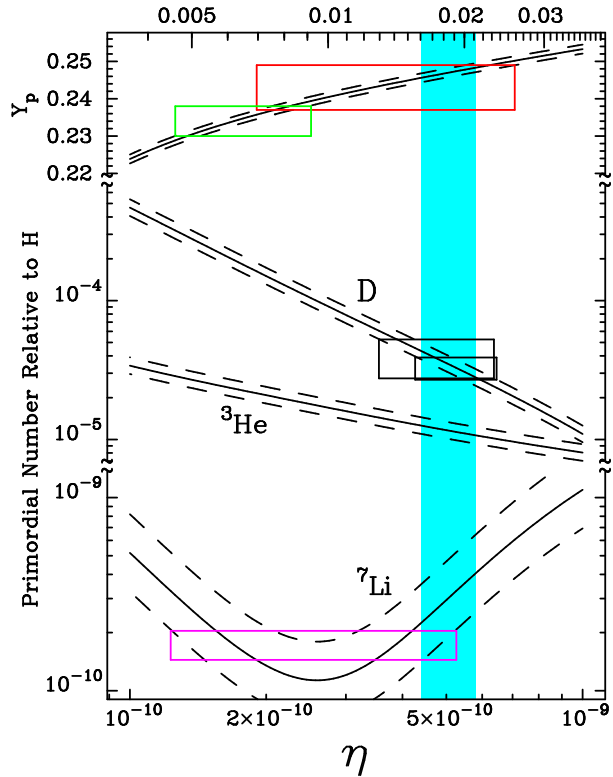,width=8cm,angle=0}}
\caption{
The predicted abundance ratios of the light elements
from Standard Big Bang Nucleosynthesis (SBBN) 
as a function of $\eta$ and $\Omega_b \, h^2$.
$^4$He is shown as primordial mass fraction, Y$_p$.  Boxes represent
95\% confidence levels of recent observational determinations.
The width of the boxes include 95\% confidence levels
in the SBBN calculations.
From Burles \& Tytler (1998b).}
\end{figure}

The high redshift quasar measurements reflect D/H as it was 
about 10-20 Gyr ago, and can
be compared to the pre-solar value  4.5 Gyr ago measured in 
terrestrial sea water,
the solar wind, comets and the atmospheres of Jupiter and Saturn 
(by measuring HD or deuterated methane in the mid-IR).  
The value of D/H today can be measured
by measuring the Lyman series in the Milky Way interstellar medium.
Spectra from $FUSE$ will be important to get the whole Lyman series,
but the main limitation of the the Milky Way measurements is the estimation
of the stellar atmosphere Lyman series absorption from which the ISM 
absorption must be deblended.  The various measurements of the
D/H ratio have been reviewed by Lemoine et al. (1999).   
Indeed, within the major uncertainties of all the measurements, the D/H
ratio appears to be declining with cosmic time, as predicted (Figure 27).

\begin{figure}[htb]
\centerline{\psfig{figure=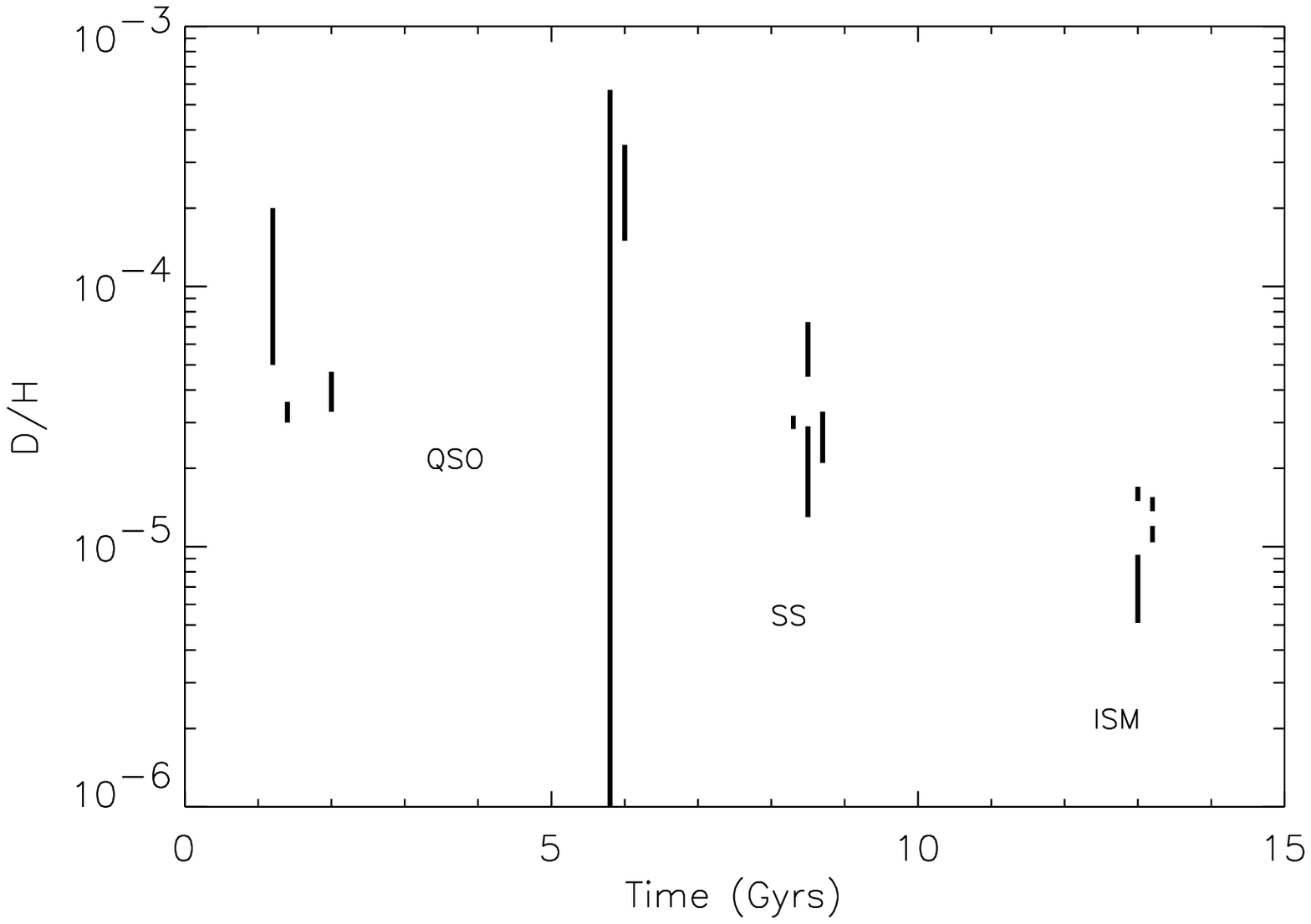,width=10cm,angle=0}}
\caption{Deuterium to hydrogen ratio as a function of cosmic
time (for $\Omega _ 0 = 1, q_o = 0.5, H_o = 50$ km/s/Mpc).  
Quasar absorbers are from Burles \& Tytler 1998ab, Songaila (1998), Webb et al. (1997), Tytler et al. (1999).  Presolar values are (SS) from the
solar wind and Jupiter.  Interstellar (ISM) values are 
from observations of stars in the Milky Way -- for complete references, 
see Lemoine et al. (1999).}
\end{figure}

\subsection{Redshifted 21-cm absorption}

A small subset of UV discovered systems show absorption in the 21-cm 
transition of neutral hydrogen (Roberts et al. 1976; Wolfe \& Davis 1979; 
Wolfe \& Briggs 1981; Briggs \& Wolfe 1983;
Briggs 1988; Taramopoulos et al. 1995; 
Carilli et al. 1996, 1997; Lane et al. 1998; Kanekar \& Chengalur 2001).  
In an elegant experiment, Briggs et al. 1989 compared the 21-cm
absorption along an extended radio source associated with a damped
Ly$\alpha$ absorber at $z\sim2$ and concluded that the absorber
has a large, disk-like geometry.  
Comparison of the 21-cm absorption with 
the column density of $H~I$ measured by the 
damped Ly$\alpha$ profile can measure the spin temperature of 
the absorbing gas (Lane et al. 1998 and references therein), 
which reflects some average
temperature along the line of sight.  The quasar absorbers seem
to have spin temperatures which are larger than those seen in
Milky Way clouds, suggesting warmer temperatures for the cold
phase of the ISM at higher redshift (Wolfe et al. 1985; 
Lane et al. 1998). 

\section{Imaging of QSO Absorbers}

They say a picture is worth a thousand words; for some, 
a picture is worth {\it ten} thousand echelle spectra.  
Despite the wealth of detailed information
derived from quasar absorption line spectra, some feel 
that one won't understand
what the absorbers $``$are" until you have a picture of them, and connect
the absorbers to classes of known objects.

\subsection{$z<1$ Mg II Absorbers}

Bergeron \& Boisse (1991) and Steidel and collaborators (Steidel,
Dickinson \& Persson 1994;
Steidel, Pettini, Dickinson \& Persson 1994; 
Steidel 1995; Steidel, Bowen, Blades
\& Dickinson 1995; Steidel et al. 1997)
undertook comprehensive imaging and spectroscopic surveys 
in order to identify the galaxies responsible 
for low redshift absorbers, primarily Mg II absorbers.
The general result was that the absorbing galaxies are drawn from 
the same population of normal galaxies seen in galaxy redshift surveys,
with relatively little evolution in color or luminosity.  Deep imaging
(Figure 28) showed that some absorbers appear to arise in the halos
of disk galaxies, supporting the result from statistical studies
that the cross-section for absorption is large. Some galaxies do 
not produce absorption in background quasars at all.    
 
\begin{figure}
\centerline{\psfig{figure=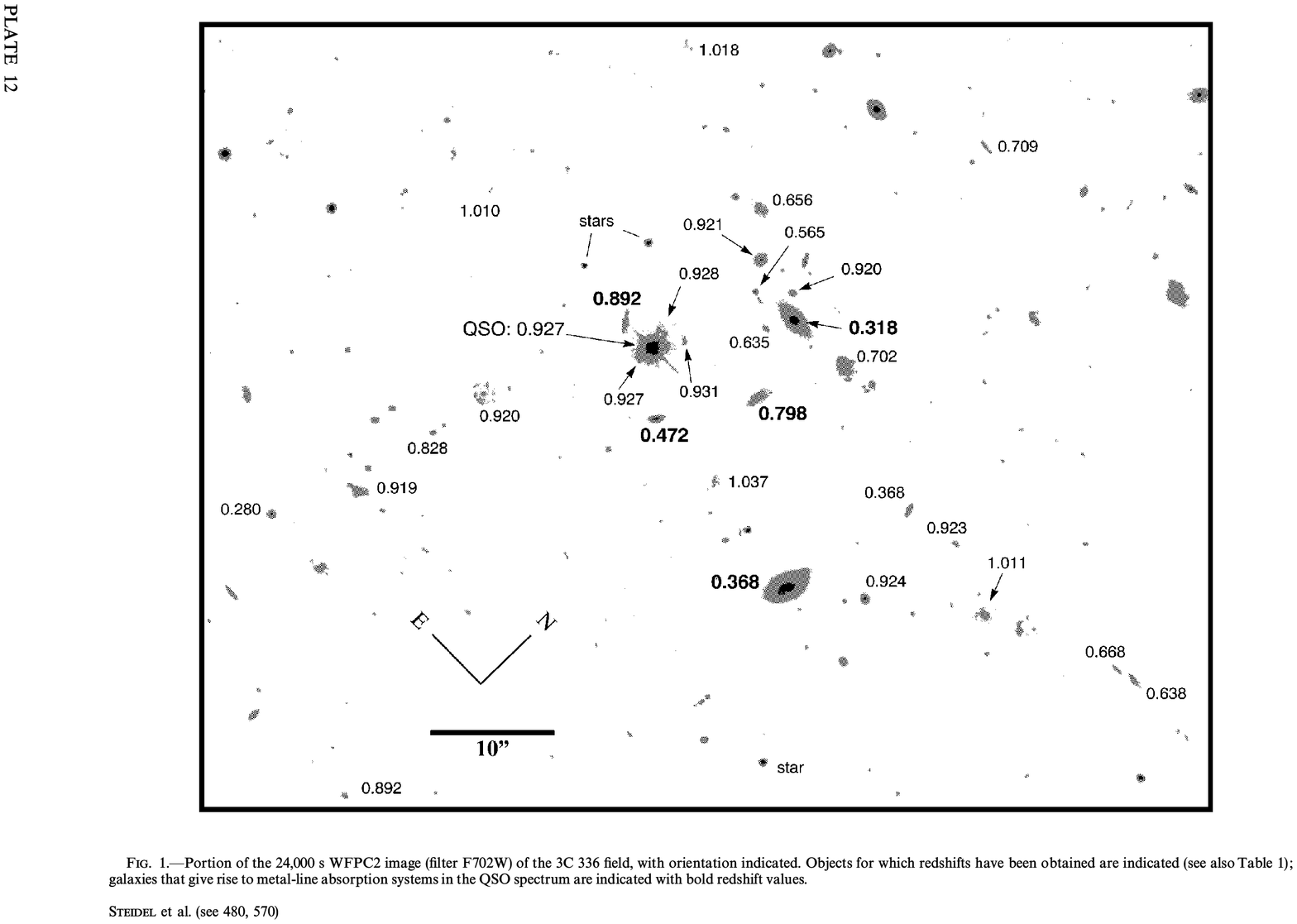,width=15cm,angle=-90}}
\caption{Deep WFPC2 image of the galaxies around quasar 
3C 336.  The bold numbers are redshifts of indicated galaxies
which show absorption in the background quasar;  the others are
redshifts of galaxies which do not show absorption.
From Steidel et al. (1997).}
\end{figure}

With few exceptions, galaxies have been found at the redshift of 
intervening Mg II absorbers, supporting the interpretation that the
absorbers are interstellar clouds of intervening galaxies.
One suggestion has been made that large absorption cross-section
is actually the result of the fact that absorbers often arise in
dwarf irregular satellites of large galaxies (York, Dopita, Green 
\& Bechtold 1984), 
analogues of the Large and Small Magellanic Clouds of the Milky Way.
Small, faint galaxies with small separations from the quasar 
image would presumably be difficult to detect. 
This is not supported by an apparent anti-correlation of Mg II equivalent
widths and increasing projected distance from the identified galaxy 
center (Steidel et al. 1994; Lanzetta et al. 1995).   

\subsection{Imaging of the Damped Lyman Alpha Absorbers} 

Deep galaxy redshift surveys typically find galaxies only to $z\sim$1.5,
so the galactic counterparts of high redshift damped Ly$\alpha$ absorbers
are searched for by means of line emission from the H II regions 
associated with them, or Lyman break techniques.
A great deal of effort went into searches for Ly$\alpha$ emission 
associated with damped Ly$\alpha$ absorbers at $z=2-3$ with limited  
success 
(Hunstead, Fletcher \& Pettini 1990;
Lowenthal et al. 1990, 1991, 1995;
Wolfe, Lanzetta, Turnshek \& Oke 1992;
Moller \& Warren 1993; 
Giavalisco, Macchetto \& Sparks 1994;
Warren \& Moller 1996;
Djorgovski et al. 1996;
Fynbo, Moller \& Warren 1999).
In many cases, 
Ly$\alpha$ may be multiply scattered by dust (Charlot \& Fall 1991, 1993).
 
Infrared surveys have been more successful in identifying damped Ly$\alpha$
absorbers via direct imaging and H$\alpha$ emission (Elston et al. 1991;
Aragon-Salamanca, Ellis \& O'Brien 1996;
Bechtold et al. 1998; 
Mannucci, Thompson, Beckwith \& Williger 1998;
Kulkarni et al. 2000, 2001;
Kobulnicki \& Koo 2000;
Warren, Moller, Fall \& Jakobsen 2001).  Wide field
infrared surveys show that the absorbers can be identified with the
same population of small, star-forming galactic fragments found
by other techniques, described in detail at this school by Marc Dickinson; 
see also Pettini et al. (2001).
Figure 29  shows the simulation of what one expects.  The bright clumps
at the intersection of the filaments 
are the regions with high enough column to produce damped Ly$\alpha$
absorption, and these probably produce stars as well.   

\begin{figure}
\caption{Distribution of neutral hydrogen 
from simulations. From Steinmetz (2000).}
\end{figure}

\section{Future Prospects} 
 
As this review was being written, a number of new facilities 
were just beginning to produce results which will enhance our 
knowledge of quasar absorbers.  $Chandra$ and $XMM$ are showing
a surprising wealth of absorption line spectral features in Seyferts and
low redshift quasars.  $FUSE$ is allowing the spectral region  
blueward of rest Ly$\alpha$ to be probed with great sensitivity,
and large surveys of absorbers at echelle resolution 
are underway with STIS on HST.
UVES on the VLT is being used to study quasars in the southern sky.
The $Cosmic$  $Origins$ $Spectrograph$ 
is planned to be installed on HST in a few years, and one
of its main science drivers is the study of quasar absorbers, particularly
He II $\lambda$304.  The Arecibo upgrade and completion of the Greenbank
Telescope will
enable more sensitive radio observations of redshifted absorption.
The Sloan Digital Sky Survey is beginning to produce new bright quasars for 
absorption line studies.  Future CMB observations with MAP and Planck 
will probably identify the redshift of HI reionization, so that searches 
may target the ionizing objects effectively.    
 
On the longer timescale, plans are  
being made to build a new generation of 30m-class ground-based telescopes, 
to be huge light-buckets for high dispersion spectroscopy.  
The $Next$ $Generation$ $Space$ $Telescope$ promises to be an 
important tool for studying the first generation of
objects responsible for reionization of the intergalactic
medium.  A large, ultraviolet optimized telescope in space
is probably the ultimate dream of workers in this field, which will 
allow high quality spectroscopy of fainter objects than are reachable
with $HST$.  Figure 30 shows a simulation of spectra expected 
with ST2010, or SUVO, a 6-8m class telescope, 
described by Morse, Shull \& Kinney (1999).     

\begin{figure}[htb]
\centerline{\psfig{figure=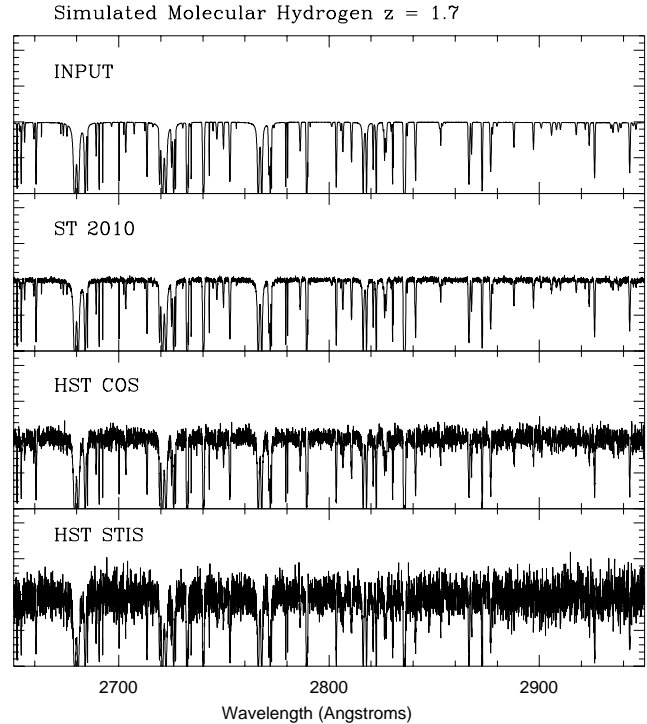,width=10cm,angle=0}}
\caption{Simulated spectrum of molecular hydrogen and the Ly$\alpha$ 
forest at $z=1.7$ as observed in 30 $HST$ orbits with STIS, 
COS and a future mission, ST2010 or SUVO.  From Bechtold (1999).}
\end{figure}

\begin{acknowledgments}
It is a pleasure to thank Ismael Perez, Marc Balcells, 
and the Director and staff of the IAC for their invitation
to participate in the Winter School, and their gracious
hospitality in Tenerife.  I thank my collaborators whose
work is quoted here in advance of publication, and longtime advice
and support from W. Sargent, S. Shectman, 
A. Wolfe, M. Shull, and P. Jakobsen.   Without Dr. Allison Stopeck
I would not have participated in the Winter school. 
Support for the preparation of this review was provided in
part by NSF grant AST-9617060. 
\end{acknowledgments}

\end{document}